\documentclass[twocolumn]{autart}    

\usepackage{graphicx}          
                               
\usepackage{multirow} 

\usepackage{enumitem} 
\usepackage{bbm} 
\usepackage{color}   
\usepackage{amsmath} 
\usepackage{amssymb}  
\usepackage{amsfonts}
\usepackage{arydshln} 

\newcommand{\vect}[1]{\boldsymbol{#1}}
\newcommand{\mat}[1]{\boldsymbol{#1}}

\newcommand{\m}[1]{\boldsymbol{#1}}
\newcommand{\mc}[1]{\mathcal{#1}}

\DeclareMathOperator{\diag}{\text{diag}}

\newtheorem{remark}{Remark}
\newtheorem{theorem}{Theorem}
\newtheorem{lemma}{Lemma}

\newtheorem{corollary}{Corollary}

\newtheorem{assumption}{Assumption}

\begin{document}

\begin{frontmatter}

\title{Continuous-time Opinion Dynamics on Multiple Interdependent Topics\thanksref{footnoteinfo}} 

\thanks[footnoteinfo]{This paper was not presented at any IFAC 
meeting. Corresponding author: M. Ye.}

\author[gro]{Mengbin Ye}\ead{mengbin.ye@anu.edu.au},    
\author[hust]{Minh Hoang Trinh}
\author[gntech]{YoungHun Lim}
\author[anu,hdu,data61]{Brian D.O. Anderson}
\author[gist]{Hyo-Sung Ahn}

\address[gro]{Jan C. Willems Center for Systems and Control, Faculty of Science and Engineering, University of Groningen, Netherlands}  
\address[anu]{Research School of Electrical, Energy and Material Engineering, Australian National University, Canberra, Australia}  
\address[hust]{School of Electrical Engineering, Hanoi University of Science and Technology (HUST), Hanoi, Vietnam}                                            
\address[gist]{School of Mechanical Engineering, Gwangju Institute of Science and Technology (GIST), Gwangju, Republic of Korea}             
\address[gntech]{Department of Electronic Engineering, Gyeongnam National University of Science and Technology, Jinju, Republic of Korea}       
\address[hdu]{School of Automation, Hangzhou Dianzi University, Hangzhou 310018, China} 
\address[data61]{Data61-CSIRO, Canberra, A.C.T. 2601, Australia} 
          
\begin{keyword}                           
opinion dynamics; social network analysis; networked systems; agent-based model               
\end{keyword}                             

\begin{abstract}                          
In this paper, and inspired by the recent discrete-time model in \cite{parsegov2017_multiissue,friedkin2016network_science}, we study two continuous-time opinion dynamics models (Model 1 and Model 2) where the individuals discuss opinions on multiple logically interdependent topics. The logical interdependence between the different topics is captured by a ``logic'' matrix, which is distinct from the Laplacian matrix capturing interactions between individuals. For each of Model 1 and Model 2, we obtain a necessary and sufficient condition for the network to reach to a consensus on each separate topic. The condition on Model 1 involves a combination of the eigenvalues of the logic matrix and Laplacian matrix, whereas the condition on Model 2 requires only separate conditions on the logic matrix and Laplacian matrix. Further investigations of Model 1 yields two sufficient conditions for consensus, and allow us to conclude that one way to guarantee a consensus is to reduce the rate of interaction between individuals exchanging opinions. By placing further restrictions on the logic matrix, we also establish a set of Laplacian matrices which guarantee consensus for Model 1. The two models are also expanded to include stubborn individuals, who remain attached to their initial opinions. Sufficient conditions are obtained for guaranteeing convergence of the opinion dynamics system, with the final opinions generally being at a persistent disagreement. Simulations are provided to illustrate the results.
\end{abstract}

\end{frontmatter}

\section{Introduction}
Recently, the study of ``opinion dynamics'' has been of particular interest to the control systems community, in part due to the similarities and parallels with multi-agent systems. The key problems involve study of models in which individuals interact and discuss opinions on a topic or set of topics, with each individual's opinion evolution described by an update rule. 

In order to understand our contribution in context, we first review several of the widely studied and most relevant models, and refer readers to the survey \cite{proskurnikov2017tutorial} for more comprehensive discussions. The French-DeGroot discrete-time model (known also as the DeGroot model) was proposed in \cite{french1956_socialpower,degroot1974OpinionDynamics}, and in it, each individual sets his/her opinion at the next time instant to be a weight average of his/her neighbours' opinions. A continuous-time counterpart was proposed in \cite{abelson1964op_dyn}. 
The Altafini model \cite{altafini2013antagonistic_interactions,proskurnikov2016opinion,liu2017altafini_exp} developed the concept that an individual may trust or distrust neighbouring individuals (captured by a positive or negative edge weight, respectively); the DeGroot and Abelson models assume individuals either trust or ignore others. 
The Friedkin-Johnsen model \cite{friedkin1990_FJsocialmodel} extended the DeGroot model to include ``stubborn'' individuals who remain somewhat attached to their initial opinion. The continuous-time counterpart of the Friedkin-Johnsen model was first proposed in \cite{taylor1968stubborn} (in fact appearing earlier than the Friedkin-Johnsen model). Decades later, the model in \cite{taylor1968stubborn} was studied as an algorithm for containment control for autonomous vehicle formations \cite{cao2012containment_SI_journal}. 

A natural extension to the above works is to consider the \emph{simultaneous} discussion of multiple topics. If the topics are independent of each other then the models discussed above may be easily extended with introduction of a Kronecker product. However, it is more likely that an individual's opinion on Topic A is influenced by his/her opinion on Topic B and vice versa. In such situations, an individual applies an introspective (internal) cognitive process to ensure that his/her opinions on all topics are logically consistent. The logical interdependences and the introspective process form the individual's \emph{belief system}. The recent works \cite{parsegov2017_multiissue,friedkin2016network_science} combined opinion evolution (as captured by the Friedkin--Johnsen opinion dynamics model) due to interaction among individuals with belief system dynamics. Since each individual's opinion set is now affected through his or her internal belief system and also the neighbours' opinions, the question of when consensus occurs becomes nontrivial, as the two processes are not guaranteed a priori to be consistent with one another. In \cite{parsegov2017_multiissue,friedkin2016network_science}, the logical interdependence is described by a matrix, and thus the model can be considered a form of \emph{matrix-weight consensus}. Matrix-weight consensus problems have recently become of interest in multi-agent systems coordination, applicable to consensus on Euclidean spaces, and bearing measurement based localisation and formation control \cite{trinh2018matrix_consensus,zhao2016localizability_bearing}.

\subsection{Contributions of This Paper}
Inspired by \cite{parsegov2017_multiissue,friedkin2016network_science}, this paper proposes and studies two continuous-time opinion dynamics models (Model 1 and Model 2) for networks of individuals simultaneously discussing logically interdependent topics. In both models, each individual is affected by three processes: (i) an introspective process using the logic matrix, common to all individuals, to secure logical consistency in the individual's opinions on the set of topics, (ii) a stubborn attachment to the individual's initial opinions, and (iii) interpersonal influence arising from sharing of opinions with neighbouring agents. The key difference between the two models proposed in the paper is in the third process, and specifically whether or not an individual assimilates the logical interdependences into his/her opinions before exchanging opinions with his/her neighbours. A discussion of the difference between the two models is postponed till their formal introduction in the next section.
	
	We begin our statement of results by obtaining separate necessary and sufficient conditions for Model 1 and Model 2 to ensure a consensus of opinions is reached when there are no stubborn individuals, i.e. individuals do not remain attached to their initial opinions. The condition on Model 1 involves \textit{a combination of the eigenvalues} of (i) the Laplacian matrix describing the network topology, and (ii) the matrix describing the logical interdependences. In contrast, the condition on Model 2 requires the eigenvalues of the two matrices to \textit{separately satisfy certain conditions}. Two sufficient conditions for consensus with no stubborn individuals, requiring only limited knowledge of the parameters of network and the individuals, are then derived for Model 1; we show that given a matrix describing the logical interdependence, one can always achieve a consensus of opinions by decreasing the strength of interactions. On the other hand, large interaction strengths sometimes results in instability. These observations on Model 1 contrast the results obtained on Model 2, and also to the discrete-time model, where in the absence of stubborn individuals, convergence of the network matrix and logic matrix separately is enough to ensure a consensus of opinions. Networks with stubborn individuals are also treated, with sufficient conditions obtained for ensuring the system is convergent for both Model 1 and Model 2. Using the obtained results, we discuss the similarities and differences of the two models, and examine the conclusions in the social context.
	
	The rest of the paper is structured as follows. Mathematical  background and two continuous-time opinion dynamics models (including a discusson of the motivations for the difference) are presented in Section~\ref{section:background_problem}.  Sections~\ref{sec:model 1} and \ref{sec:model2} study convergence conditions of the two models, respectively, with and without stubborn individuals. Simulations are provided in Section~\ref{sec:simulations}, with conclusions drawn in Section~\ref{sec:con}.

\section{Background and Formal Problem Statement}\label{section:background_problem}

We begin by introducing some mathematical notations. Let $\vect 1_n$ and $\vect 0_n$ denote, respectively, the $n\times 1$ column vectors of all ones and all zeros. For a vector $\vect x\in\mathbb{R}^n$, $0\leq\vect x$ and $0 < \vect x$ indicate component-wise inequalities, i.e., for all $i\in\{1,2,\ldots,n\}$, $0\leq x_i$ and $0<x_i$, respectively. The canonical basis of $\mathbb{R}^n$ is given by $\mathbf{e}_1, \ldots, \mathbf{e}_n$. We denote $\sqrt{-1} = \jmath$ as the imaginary unit, and for a complex number $z = a+b\jmath$ we denote $\mathfrak{Re}(z) = a$ and $\mathfrak{Im}(z) = b$. The modulus is $\vert z \vert = \sqrt{a^2+b^2}$. For a matrix $\mat{A}\in\mathbb{R}^{n\times m}$, we denote its $\infty$-norm as $\Vert \mat{A} \Vert_{\infty} = \max_{1\leq i \leq n} \sum^m_{j=1} \vert a_{ij} \vert$. The Kronecker product is given by $\otimes$. Note that the terms ``node'', and ``individual'' are used interchangeably. The following result will be used later in this paper.



\begin{theorem}[{\cite[Theorem 6.1.1]{horn2012matrixbook}}]\label{thm:gersgorin}
	Let $\mat{A} = \{a_{ij}\} \in \mathbb{R}^{n\times n}$, and let $R_i (\mat{A}) = \sum_{j=1, j\neq i}^n \vert a_{ij} \vert$ for $i = 1, \hdots, n$ denote the absolute row sums of the off-diagonal elements of $\mat{A}$. Consider the $n$ Ger\v{s}gorin discs
	\begin{equation}
	\{z \in \mathbb{C} : \vert z-a_{ii} \vert \leq R_i(\mat{A}) \},\quad i = 1, \hdots, n
	\end{equation}
	The eigenvalues of $\mat{A}$ are in the union of the Ger\v{s}gorin discs $H(\mat{A}) = \bigcup_{i=1}^n \{z \in \mathbb{C} : \vert z-a_{ii} \vert \leq R_i(\mat{A}) \}.$
	Furthermore, if the union of $k$ of the $n$ discs that comprise $H(\mat{A})$ forms a set $H_k(\mat{A})$ that is disjoint from the remaining $n-k$ discs, then $H_k(\mat{A})$ contains exactly $k$ eigenvalues of $\mat{A}$, counted according to their algebraic multiplicities.
\end{theorem}

\subsection{Graph Theory}
The interaction between $n$ individuals in a social network is modelled using a weighted directed graph, denoted as $\mathcal{G} = (\mathcal{V}, \mathcal{E}, \mathcal{A})$. Each individual is a node in the finite, nonempty set of nodes $\mathcal{V} = \{v_i : i \in \mathcal{I} = \{1, \hdots, n\}\}$. The set of ordered edges is $\mathcal{E} \subseteq \mathcal{V}\times \mathcal{V}$. We denote an ordered edge as $e_{ij} = (v_i, v_j) \in \mathcal{E}$. An edge $e_{ij}$ is said to be outgoing with respect to $v_i$ and incoming with respect to $v_j$, and connotes that individual $j$ learns of, and takes into account, the opinion value of individual $i$  when updating its own opinion. The (incoming) neighbour set of $v_i$ is defined as $\mathcal{N}_i = \{v_j \in \mathcal{V} : e_{ji} \in \mathcal{E}\}$. The weighted adjacency matrix $\mathcal{A}\in\mathbb{R}^{n\times n}$ of $\mathcal{G}$ has nonnegative elements $a_{ij}$ satisfying $a_{ij} > 0 \Leftrightarrow e_{ji}\in \mathcal{E}$, and it is assumed that $a_{ii} = 0,  \forall i$. The Laplacian matrix, $\mathcal{L} = [l_{ij}]_{n\times n}$, of the associated digraph $\mathcal{G}$ is defined as $l_{ii} = \sum^n_{k=1,k\neq i} a_{ik}$ and $l_{ij}= -a_{ij}$ for $j\neq i$.
A directed path is a sequence of edges of the form $(v_{p_1}, v_{p_2}), (v_{p_2}, v_{p_3}), ...,$ where $v_{p_i} \in \mathcal{V}, e_{p_{i}p_{i+1}} \in \mathcal{E}$. Node $i$ is reachable from node $j$ if there exists a directed path from $v_j$ to $v_i$. A node $v_i$ is called a root if there is a path from $v_i$ to every $v_j \in \mathcal{V}, j\neq i$. A directed spanning tree is a directed graph formed by directed edges of the graph that connects all the nodes, and where every vertex apart from the unique root node has exactly one parent. A graph is said to contain a directed spanning tree if a subset of the edges forms a directed spanning tree\footnote{Some literature use other terms, e.g. rooted out-branching or directed rooted tree.}. A graph is strongly connected if and only if, for every node $v_i$, there exists a directed path to every other node $v_j$. The following is a standard result that will be used throughout this paper.

\begin{lemma}[\cite{mesbahi2010graph}]\label{lem:DST_graph}
	The Laplacian $\mathcal{L}$ associated with a graph $\mathcal{G} = (\mathcal{V}, \mathcal{E}, \mathcal{A})$ has a single eigenvalue at $0$ if and only if $\mathcal{G}$ has a directed spanning tree. Associated with the single $0$ eigenvalue are left and right eigenvectors $\vect{\gamma} \geq 0$ and $\vect{1}_n$, respectively, with normalisation $\vect{\gamma}^\top\vect{1}_n = 1$. All other eigenvalues have strictly positive real part.
\end{lemma}
If the graph contains a directed spanning tree, then there exists an $r \leq n$ such that the nodes reordered $v_1, \hdots, v_r$ induce a maximally closed and strongly connected subgraph $\mathcal{G}_{L}$. By closed, we mean that no edges are incoming to $\mathcal{G}_{L}$. We denote by $\mathcal{G}_{F}$ the subgraph induced by the set of nodes $v_{r+1}, \hdots, v_n$. With the nodes reordered, the Laplacian matrix $\mathcal{L}$ associated with $\mathcal{G}$ is expressed as \begin{equation}\label{eq:laplacian_ordered}
\mathcal{L} = \begin{bmatrix} \mathcal{L}_{11} & \mat{0}_{r \times (n-r)} \\ \mathcal{L}_{21} & \mathcal{L}_{22} \end{bmatrix}.
\end{equation}
where $\mathcal{L}_{11} \in \mathbb{R}^{r\times r}$ is irreducible. If $r=n$ then $\mathcal{G}$ is strongly connected, and $\mathcal{L}_{22}$ vanishes. The matrices $\mathcal{L}$ and $\mathcal{L}_{11}$ are singular $M$-matrices \cite{xia2017ground_laplacian}, which implies in light of Lemma~\ref{lem:DST_graph} that if $r < n$ then $\mathcal{L}_{22}$ is a nonsingular $M$-matrix, i.e. its eigenvalues have positive real part.  Moreover, if $\mat D_1, \mat D_2$ are nonnegative diagonal matrices of appropriate size, and $\mat D_1$ has at least one positive diagonal entry, then all eigenvalues of $\mathcal{L}_{11} + \mat D_1$ and $\mathcal{L}_{22} + \mat D_2$ have positive real part (see \cite[Theorem 2.3]{berman1979nonnegative_matrices} and \cite[Corollary 4.33]{qu2009cooperative_book}, respectively).
The left eigenvector $\vect{\gamma}^\top = [\gamma_1, \hdots, \gamma_n]$ defined in Lemma~\ref{lem:DST_graph} has entries $\gamma_i > 0,~i \in \{1, \hdots, r\}$. 

\subsection{Opinion Dynamics Model and Problem Statement}\label{ssec:opinion_rules}
 We now present two general opinion dynamics models and the formal problem statement. We then provide details on the motivation behind the proposed models, including discussion and exploration of a key matrix describing the logical interdependence of the topics. 

Given a population of $n$ individuals, indexed by $\mathcal{I} = \{1, \hdots, n\}$, let $\vect{x}_i(t) = [x_i^1(t), \hdots, x_i^d(t)]^\top \in \mathbb{R}^{d}$ be the vector of opinion values\footnote{One illustrative example is where the $k^{th}$ entry $x_i^k$ represents individual $i$'s belief/certainty in a statement defining topic $k$ which in principle is provable to be true or false. Alternatively, $x_i^k$ may represent an attitude towards adoption of an idea defining topic $k$, with negative values (respectively positive values) representing refusal (respectively willingness) to adopt. More details are provided in Section~\ref{ssec:class_C}.} held by individual $i \in \mathcal{I}$, at time $t$, on $d$ different topics. We index the topics by $\mathcal{J} = \{1, \hdots, d\}$. Where there is no confusion, we drop the time argument $t$. We propose two models to describe how the opinions of individual $i$ evolve.
	
	\textit{Model 1:}
	\begin{align}\label{eq:xi_update_C_stub}
	\dot{\vect{x}}_i(t) &= \sum_{j\in\mathcal{N}_i} a_{ij} \mat{C} \left( \vect{x}_j(t) - \vect{x}_i(t) \right) + (\mat{C} - \mat{I}_{d})\vect{x}_i(t) \nonumber \\
	& \quad \quad + b_i (\vect{x}_i(0) - \vect{x}_i(t)).
	\end{align}
	\textit{Model 2:}
	\begin{align}\label{eq:xi_update_C_stub_R}
	\dot{\vect{x}}_i(t) &= \sum_{j\in\mathcal{N}_i} a_{ij} \left( \vect{x}_j(t) - \vect{x}_i(t) \right) + (\mat{C} - \mat{I}_{d})\vect{x}_i(t) \nonumber \\
	& \quad \quad + b_i (\vect{x}_i(0) - \vect{x}_i(t)).
	\end{align}
In the above, $a_{ij}$ is the $(i,j)^{th}$ entry of the adjacency matrix $\mathcal{A}$ associated with the graph $\mathcal{G}$. The constant matrix $\mat{C} \in \mathbb{R}^{d\times d}$, which is the same for each individual $i\in \mathcal{I}$, represents the logical interdependence/coupling between different topics. The scalar $b_i \geq 0$ is a measure of individual $i$'s stubbornness, or attachment to his/her initial opinion value $\vect{x}_i(0)$.  When $\mat{C} = \mat{I}_d$, Eq.~\eqref{eq:xi_update_C_stub} and \eqref{eq:xi_update_C_stub_R} are equivalent, and the motivations and dynamical properties for this special case are comprehensively detailed in \cite{proskurnikov2017tutorial,abelson1964op_dyn,taylor1968stubborn,cao2012containment_SI_journal}. For the general case of $\mat{C} \neq \mat{I}_d$ as investigated in this paper, the role and properties of $\mat{C}$, and the differences between \eqref{eq:xi_update_C_stub} and \eqref{eq:xi_update_C_stub_R}, are explained in Section~\ref{ssec:interdependent_topics} below, after we complete a formal introduction of the model. 

\textit{Model 1:} The dynamical system describing a network of individuals using \eqref{eq:xi_update_C_stub} can be expressed as 
	$\dot{\vect{x}} = (\mat{I}_n \otimes (\mat{C} - \mat{I}_d))\vect{x} - (\mathcal{L}\otimes \mat{C})\vect{x} + (\mat{B} \otimes \mat{I}_d)(\vect{x}(0) - \vect{x})$, 
	where $\vect{x} = [\vect{x}_1^\top, \hdots, \vect{x}_n^\top]^\top \in \mathbb{R}^{nd}$ is the stacked vector of all opinion vectors $\vect{x}_i$ and $\mathcal{L}$ is the Laplacian matrix associated with the graph $\mathcal{G}$. The diagonal matrix $\mat{B} = \diag[b_i]$ encodes individuals' stubbornness. One can rearrange to obtain
	\begin{align}\label{eq:compact_network_stubborn_02}
	\dot{\vect{x}} & = -\big(\mat{I}_{nd} + (\mathcal{L}- \mat{I}_n)\otimes \mat{C} + \mat{B} \otimes \mat{I}_d \big)\vect{x} \nonumber \\
	& \quad \qquad + (\mat{B}\otimes \mat{I}_d)\vect{x}(0).
	\end{align}
	If $b_i = 0\,\forall\,i$ then \eqref{eq:xi_update_C_stub} reduces to
	\begin{equation}\label{eq:xi_update_C}
	\dot{\vect{x}}_i(t) = \sum_{j\in\mathcal{N}_i} a_{ij} \mat{C} \left( \vect{x}_j(t) - \vect{x}_i(t) \right) + (\mat{C} - \mat{I}_d)\vect{x}_i(t),
	\end{equation}
	with the network dynamics being
	\begin{align}
	\dot{\vect{x}} = -\big( \mat{I}_{nd} + (\mathcal{L}-\mat{I}_n) \otimes \mat{C}  \big) \vect{x}. \label{eq:x_system_continuous_C_02}
	\end{align}
	
	\textit{Model 2:} Similar to the above, one can show that the network dynamics of \eqref{eq:xi_update_C_stub_R} are 
\begin{align}
\dot{\m{x}}(t) & = -\big((\mc{L}+\m{B})\otimes \mat I_d + \mat I_n\otimes(\m{I}_d - \m{C}) \big) \m{x}(t) \nonumber \\
& \quad \quad + (\m{B}\otimes \m{I}_d) \m{x} (0). \label{eq:model_stubborn}
\end{align} 
	If $b_i = 0\,\forall\,i$, \eqref{eq:xi_update_C_stub_R} becomes
	\begin{equation}\label{eq:xi_update_C_R}
	\dot{\vect{x}}_i(t) = \sum_{j\in\mathcal{N}_i} a_{ij} \left( \vect{x}_j(t) - \vect{x}_i(t) \right) + (\mat{C} - \mat{I}_d)\vect{x}_i(t)
	\end{equation}
	with the network dynamics given by
	\begin{align}\label{eq:3}
	\dot{\m{x}}(t) = - \big( \mc{L} \otimes \m{I}_d + \m{I}_n \otimes (\m{I}_d - \m{C}) \big) \m{x}(t).
	\end{align}

The problem considered in this paper is as follows. Let a social network be represented by a directed graph $\mathcal{G} = (\mathcal{V}, \mathcal{E}, \mathcal{A})$. Supposing that all individuals either use the opinion updating rule \eqref{eq:xi_update_C_stub} or \eqref{eq:xi_update_C_stub_R}, we seek to determine $1)$ the connectivity conditions (including constraints on the edge weights) on the graph $\mathcal{G}$, $2)$ the conditions on the matrix $\mat{C}$, and $3)$ conditions on $\mat{B}$, which guarantee that as $t\to \infty$, opinions reach a steady value, i.e. $\dot{\vect{x}}_i = 0, \forall\,i\in\mathcal{I}$. A special case of convergence is consensus of opinions\footnote{Some literature on consensus and containment control study agents with dynamics $\dot{\vect{x}}_i = \mat{F}\vect{x}_i + \mat{G}\vect{u}_i$, with control $\vect{u}_i = \mat{K} \sum_{j\in\mathcal{N}_i} a_{ij} \left( \vect{x}_j(t) - \vect{x}_i(t) \right)$. A typical result requires the control gain $\mat{K} = \mat{G}^\top \mat{Z}$ where $\mat{Z} = \mat{Z}^\top > 0$ is the solution to some algebraic Riccati equation, see e.g. \cite{qin2015collective_linear_consensus}. The eigenvalues of $\mat{GG}^\top\mat{Z}$ are therefore nonnegative and real \cite[Corollary 7.6.2]{horn2012matrixbook}. In \eqref{eq:xi_update_C_stub}, $\mat{C} - \mat{I}_{d}$ and $\mat{C}$ replaces $\mat{F}$ and $\mat{GG}^\top \mat{Z}$, respectively. We allow $\mat{C}$ to have complex eigenvalues, which increases the number of different interdependencies between topics describable by $\mat{C}$. Moreover, $\mat{F}$ is also a function of $\mat{C}$, which means conditions for convergence are different to existing results.}. We say that a consensus on opinions has been reached if
 \begin{align}\label{eq:consensus_cond}
\lim_{t\to\infty} \Vert \vect{x}_i - \vect{x}_j \Vert = 0\,,~ \forall\, i,j \in \mathcal{I}.
\end{align}
It will be shown that consensus can occur when there are no stubborn individuals in the network, or if there exist stubborn individuals and $\vect{x}_i(0) = \vect{x}_j(0),\forall\,i,j$
.

\begin{remark}
The condition in \eqref{eq:consensus_cond} holds when a consensus of opinions is reached for every topic, i.e. $\lim_{t\to\infty} x_i^p(t) = \lim_{t\to\infty} x_j^p(t), \forall\, i,j \in \mathcal{I}$ and $p\in \mathcal{J}$. Part of this paper focuses on establishing conditions for which \eqref{eq:consensus_cond} is achieved when there are no stubborn individuals in the network. In our recent paper \cite{ye2018_CDC_logic}, we study the discrete-time version of \eqref{eq:xi_update_C}, and identify that partial consensus can arise when certain conditions on $\mat{C}$ are met. By partial consensus, we mean that consensus occurs for a subset of topics only: $\lim_{t\to\infty} x_i^p = \lim_{t\to\infty} x_j^p$ for all $i,j\in \mathcal{I}$ while $\lim_{t\to\infty} x_i^q = \lim_{t\to\infty} x_j^q, i,j \in \mathcal{I}$, for some $p, q \in \mathcal{J}$ and $p\neq q$. In particular, a necessary condition is that there is heterogeneity of the logic matrix between individuals. In future works, we aim to conduct similar investigations into conditions for partial consensus for the continuous-time model proposed in this paper.
\end{remark}

 \begin{remark}
 A generalisation of \eqref{eq:xi_update_C_stub} where individual $i$ remains attached to several other static opinions, in addition to $i$'s initial opinion $\vect{x}_i(0)$, is given in \cite{taylor1968stubborn}. Suppose that individual $i$ considers $m$ different constant inputs $\vect{u}_{i,1}, \hdots, \vect{u}_{i,m}$, such as the initial opinions of his/her neighbours, or constant information sources from the media. Then, the last term in \eqref{eq:xi_update_C_stub} becomes $\sum_{k=1}^m b_{i,k} (\vect{u}_{i,k} - \vect{x}_i(t)) = \bar{b}_i(\bar{\vect u}_i - \vect{x}_i(t))$ where $\bar{b}_i = \sum_{k=1}^m b_{i,k}$ and $\bar{\vect u}_i = \frac{1}{\sum_{k=1}^m b_{i,k}} \sum_{k=1}^m b_{i,k} \vect{u}_{i,k}$ is the aggregate influence of all external influences. This paper focuses on convergence analysis, results which hold for both \eqref{eq:xi_update_C_stub} and the generalised model in \cite{taylor1968stubborn}. Future research will focus on how $\vect{u}_{i,1}, \hdots, \vect{u}_{i,m}$ affect the final opinion distribution $\vect{x}(\infty)$.
 \end{remark}

\subsection{Interdependent Topics and the $\mat{C}$ Matrix}\label{ssec:interdependent_topics}
The concept of an opinion dynamics model for capturing simultaneous discussion on multiple logically interdependent topics was first proposed in discrete-time \cite{parsegov2017_multiissue,friedkin2016network_science}. In \cite{parsegov2017_multiissue,friedkin2016network_science}, the authors capture this with a matrix of \emph{multi-issues dependence structure} (MiDS). Similarly, we define in this paper a \emph{logic matrix}  $\mat{C}$ which encodes the logical coupling between issues, which has some different properties to MiDS matrix in \cite{parsegov2017_multiissue,friedkin2016network_science}. We now provide an example to motivate $\mat{C}$ and demonstrate its purpose in a person's cognitive process for handling logically interdependent topics. 

Consider two topics being simultaneously discussed; $1)$ mentally challenging tasks are just as exhausting as physically challenging tasks and $2)$ that chess should be considered a sport in the Olympics. Let individual $i$'s opinion vector be $\vect{x}_i = [x^1_i, x^2_i]^\top$. For topic $1$, if $x_i^1$ is positive (respectively negative) then individual $i$ believes mentally challenging tasks are just as exhausting (respectively not as exhausting) as physically challenging tasks. For topic $2$, if $x_i^2$ is positive (respectively negative) then individual $i$ believes chess should be considered (respectively not considered) an Olympic sport. One possible logic matrix is given by
\begin{equation}\label{eq:C_example}
\mat{C} = \begin{bmatrix}
1 &  0 \\ 0.7 & 0.3
\end{bmatrix}
\end{equation}
which indicates that individual $i$ believes whether an event should be in the Olympics depends heavily on whether it is exhausting. While the above $\mat{C}$ is row-stochastic, we do not in general require $\mat{C}$ to be row-stochastic (though other constraints will apply).

To gain further insight into constraints on $\mat{C}$ and each individual's internal process for securing logical consistency, and by way of example, suppose that $\mathcal{N}_i = \{ \emptyset \}$ and $b_i = 0$. Then \eqref{eq:xi_update_C_stub} and \eqref{eq:xi_update_C_stub_R} become
\begin{equation}\label{eq:xi_cognitiveprocess}
\dot{\vect{x}}_i = \mat{C}\vect{x}_i - \vect{x}_i
\end{equation}
Here, the matrix $\mat{C}$ is the logic matrix detailed in Section~\ref{ssec:interdependent_topics}, and $(\mat{C}-\mat{I}_d)\vect{x}_i$ \emph{is the difference} between individual $i$'s current opinion $\vect{x}_i$ and its opinions \textit{after assimilating} the logical interdependencies of the discussed topics, $\mat{C}\vect{x}_i$.  Existing literature indicates that individuals will use an \emph{introspective (internal) cognitive process} to remove cognitive inconsistencies in their set of beliefs \cite{festinger1962cognitive_dissonance,gawronski2012cognitive_book,converse1964beliefsystem}, and this process is represented in individual $i$ by the dynamics of \eqref{eq:xi_cognitiveprocess}. 

Returning to the example of chess and Olympic sports, suppose that individual $i$ has initial opinions $\vect{x}_i(0) = [1, -1]^\top$. Then \eqref{eq:xi_cognitiveprocess} with $\mat{C}$ given in \eqref{eq:C_example} yields $\lim_{t\to\infty} \vect{x}_i = [1,1]^\top$. In other words, individual $i$ has an initial opinion against chess being an Olympic sport, but his/her logical reasoning that mentally challenging tasks are just as exhausting creates an \emph{inconsistency}. Individual $i$ uses a cognitive process, viz. \eqref{eq:xi_cognitiveprocess}, to adjust his/her opinions until a consistent set of opinions is held.

The fact that \eqref{eq:xi_cognitiveprocess} represents a cognitive process implies that some constraints must be placed on $\mat{C}$. We assume that \eqref{eq:xi_cognitiveprocess} will eventually lead to a \emph{consistent} belief system. We therefore do not expect $\vect{x}_i(t)$ to oscillate indefinitely, or for $\lim_{t\to\infty} \Vert \vect{x}_i(t) \Vert = \infty$. If \eqref{eq:xi_cognitiveprocess} is asymptotically stable then $\lim_{t\to\infty} \vect{x}_i = \vect{0}_d$, which is a non-generic cognitive process, and we therefore assume does not occur. Thus, one expects in general that $\lim_{t\to\infty} \vect{x}(t)$ exists under \eqref{eq:xi_cognitiveprocess} and is nonzero. In order for \eqref{eq:xi_cognitiveprocess} to have these properties, we impose the following assumption.
	\begin{assumption}\label{assm:C_constant}
		The matrix $\mat{C}$, with eigenvalues $\lambda_k(\mat{C})$, has a semi-simple\footnote{By semi-simple, we mean that the geometric and algebraic multiplicities are the same. Equivalently, the Jordan blocks of the eigenvalue $1$ are all 1 by 1.} eigenvalue at 1 with multiplicity $p \geq 1$, ordered as $\lambda_1(\mat{C}) = \hdots = \lambda_p(\mat{C}) = 1$, with associated right and left eigenvectors $\vect{\zeta}_r$ and $\vect{\xi}_r^\top$), respectively, satisfying $\vect{\xi}_r^\top \vect{\zeta}_r = 1$ for $r = 1, \hdots, p$. Other eigenvalues $\lambda_k(\mat{C})$ satisfy $\mathfrak{Re}(\lambda_k(\mat C)) < 1, \forall\, k > p$, and $c_{ii} \geq 0\, \forall\,i \in \mathcal{J}$.
	\end{assumption}
	The eigenvalue assumptions are necessary and sufficient for \eqref{eq:xi_cognitiveprocess} to have the desired convergence properties. The requirement that $c_{ii} \geq 0$ for all $i \in \mathcal{J}$ simply (and reasonably) indicates that topic $i$ is nonnegatively coupled to itself. No restrictions are placed on the off-diagonal entries of $\mat{C}$, i.e. how two different topics are coupled. A special case of Assumption~\ref{assm:C_constant} is $\mat{C} = \mat{I}_d$. 
	We note here that $\vect{\zeta}_i, i = 1, \hdots, p$ is a nullvector of $\mat{C}-\mat{I}_d$, and the introspective dynamics \eqref{eq:xi_cognitiveprocess} will yield that $\vect{x}_i(\infty)$ is in the span of $\{\vect{\zeta}_r \}, r = 1, \hdots, p$. It will become apparent that $\vect{\xi}_r, \vect{\zeta}_r, r = 1, \hdots, p$ also play a role in determining the final set of opinions for the network of individuals.

Next, one may naturally ask whether, with $b_i > 0$, the introspective process varies from \eqref{eq:xi_cognitiveprocess} to become
\begin{align}\label{eq:xi_cognitive_stub_false}
\dot{\vect{x}}_i(t) = (\mat{C} - \mat{I}_d)\vect{x}_i(t) +  b_i\mat{C} (\vect{x}_i(0) - \vect{x}_i(t))
\end{align}
or the second summand has no logic matrix $\mat{C}$, i.e.
\begin{align}\label{eq:xi_cognitive_stub_true}
\dot{\vect{x}}_i(t) = (\mat{C} - \mat{I}_d)\vect{x}_i(t) +  b_i (\vect{x}_i(0) - \vect{x}_i(t)).
\end{align}
The latter is our proposed model. Suppose that Assumption~\ref{assm:C_constant} holds. For large $b_i$, \eqref{eq:xi_cognitive_stub_false} can become unstable but \eqref{eq:xi_cognitive_stub_true} remains convergent. We argue that larger values of stubbornness in an individual should not create an unstable belief system, and therefore \eqref{eq:xi_cognitive_stub_true} represents a stubborn individual's introspective cognitive process. 

Last, we examine the differences in the first terms of \eqref{eq:xi_update_C_stub} and \eqref{eq:xi_update_C_stub_R}, which capture the interpersonal interactions.
	
	\textit{Model 1:} The first term of \eqref{eq:xi_update_C_stub} describes that individual $i$ displays his/her own opinions \textit{after} assimilating the logical interdependencies using $\mat C$, i.e. $\mat C \vect x_i$ is displayed, and learns of neighbour $j$'s already assimilated opinions, $\mat C\vect x_j(t)$. Individual $i$'s rate of opinion change is then influenced by the weighted difference in assimilated (displayed) opinions between himself/herself, and his/her neighbours, $\sum_{j\in \mathcal{N}_i}a_{ij} (C\vect x_j(t)-\mat C\vect x_i(t))$. Thus, \eqref{eq:xi_update_C_stub} captures the simultaneous effect of three different processes, viz. (i) interpersonal influence due to differences in \textit{assimilated opinions} between individual $i$ and neighbour individuals $j$, (ii) an introspective cognitive process for securing logical consistency between topics, and (iii) a stubborn attachment to $i$'s initial prejudices/opinions.

	\textit{Model 2:} In contrast, the first term of \eqref{eq:xi_update_C_stub_R} reflects that individual $i$ displays opinions $\vect x_i(t)$ without assimilation, then learns of opinions $\vect x_j(t)$ also without assimilation. Individual $i$'s rate of opinion change is influenced by the weighted difference in unassimilated (displayed) opinions between himself/herself and his/her neighbours, $\sum_{j\in \mathcal{N}_i} a_{ij} (\vect x_j(t)- \vect x_i(t))$. Thus, in \eqref{eq:xi_update_C_stub_R}, individual $i$ only uses the introspective process (second term) to internally assimilate the logical interdependencies into his/her opinions.
	
	The models in this paper are inspired by the discrete-time model in \cite{parsegov2017_multiissue,friedkin2016network_science}. In \cite[Supplementary Material, Remark 1]{friedkin2016network_science}, the authors describe two variations for the discrete-time opinion dynamics when multiple logically interdependent topics are simultaneously discussed. In discrete-time, the two variations yield identical difference equations when $\mat C$ is homogeneous among the individuals, and no variation candidate is stated as being clearly more accepted. Thus, we developed both variations in continuous-time to obtain Models 1 and 2 as alternative models for continuous-time opinion evolution for logically interdependent topics. We study both in order to better understand the dynamics of the two processes, including any differences. It turns out that whether individuals exchange assimilated opinions (Model 1) or unassimilated opinions (Model 2) can lead to different convergence and stability properties.
	
	\begin{remark}
		We wish to clarify that $\mat {Cx}_i(t)$ represents individual $i$ using $\mat C$ to \textit{assimilate} the logical interdependences into an opinion vector $\vect x_i(t)$ to obtain an opinion vector $\mat {Cx}_i(t)$. On the other hand \eqref{eq:xi_cognitiveprocess} is the proposed model of the introspective process by which individual $i$ ensures that he or she eventually has a set of opinions which are consistent with the logical interdependence structure, i.e. a consistent belief system. Such a model guarantees that $\vect x_i(\infty)$ is a fixed point of the linear map $\mat C$; $\mat C\vect x_i(\infty) = \vect x_i(\infty)$.
	\end{remark}
	
	Last, we state an assumption on the graph $\mathcal{G}$ representing the interpersonal interaction topology of the network.
	
	\begin{assumption}\label{assm:graph}
		The graph $\mathcal{G} = (\mathcal{V}, \mathcal{E}, \mathcal{A})$ has a directed spanning tree, and the nodes are ordered such that the associated Laplacian matrix $\mathcal{L}$ takes the form of \eqref{eq:laplacian_ordered}.
	\end{assumption}

If $\mathcal{G}$ does not have a directed spanning tree, then there exist at least two closed and strongly connected subgraphs, with each subgraph containing at least one individual. \textit{Regardless of whether Model 1 or Model 2} is used to capture the opinion dynamics, the opinions of the individual(s) in each closed subgraph evolve independently of the opinions of other individuals, and for almost all $\vect x(0)$, will not not reach a consensus with the opinions of any other individual in the network. By ``almost all $\vect x(0)$'', it is meant that there may be a proper subset $\mathcal{F}$ of $\mathbb{R}^{nd}$ with Lebesgue measure zero, for which consensus can still be reached if $\vect x(0) \in \mathcal{F}$. This implies that for almost all $\vect x(0)$, $\mathcal{G}$ having a directed spanning tree (Assumption~\ref{assm:graph}) is a \textit{necessary condition} for consensus to be achieved.  
	
	We present the convergence analysis in the following two sections, and defer discussion of the results, including comparison between the two models, until the end.


\section{Networks of Individuals with Model 1}\label{sec:model 1}
In this section, we investigate the convergence properties of networks where each individual's opinion vector have dynamics given by \eqref{eq:xi_update_C_stub}.

\subsection{Consensus With No Stubborn Individuals}
We first present the main convergence result when there are no stubborn individuals, i.e. $b_i = 0, \forall\,i \in\mathcal{I}$. 

\begin{theorem}\label{thm:stability_continuous_C}
	Let $\mat{C}$, which satisfies Assumption~\ref{assm:C_constant}, and $\mathcal{G} = (\mathcal{V},\mathcal{E},\mathcal{A})$ be given, with $\lambda_i (\mathcal{L})$ and $\lambda_k(\mat{C})$ being the eigenvalues of the Laplacian matrix $\mathcal{L}$ and logic matrix $\mat{C}$, respectively. The eigenvalues are ordered such that $\lambda_1(\mathcal{L}) = 0$, and $\lambda_1(\mat{C}), \hdots,  \lambda_p(\mat{C})$ are the $p\geq 1$ semi-simple eigenvalues at 1.
	
	Then, with each individuals' opinions evolving according to \eqref{eq:xi_update_C}, the social network reaches a consensus on all topics exponentially fast if and only if 
	\begin{equation}	\label{cond:stable_C_02} 
	\mathfrak{Re} \left( ( 1 - \lambda_i(\mathcal{L}) ) \lambda_k ( \mat{C} ) \right) < 1, \; \forall\, i \in \mathcal{I}\setminus \{1\}\; \mathrm{ and }\; k \in \mathcal{J},
	\end{equation} 
	Moreover, with $\vect{\gamma}^\top$ as defined in Assumption~\ref{assm:graph}, and $\vect{\xi}_r^\top$ and $\vect{\zeta}_r$ as defined in \ref{assm:C_constant}, the solution satisfies 
	\begin{align}\label{eq:stability_continuous_C_solution}
	\lim_{t\to\infty} \vect{x}_i (t) = \Big(\sum_{r=1}^p \vect{\zeta}_r \vect{\xi}_r^\top \Big) \sum_{j=1}^n \gamma_j \m{x}_j(0), \forall i \in \mathcal{I},
	\end{align}
\end{theorem}
\emph{\textbf{Proof}:} Observe that \eqref{cond:stable_C_02} holds only if $\lambda_i (\mathcal{L}) \neq 0$, $i = 2,...,n$, which in turn holds if and only if $\mathcal{G}$ has a directed spanning tree (see Lemma~\ref{lem:DST_graph}).

We first establish the sufficiency of \eqref{cond:stable_C_02}. With $b_i = 0, \forall i \in \mathcal{I}$, the opinions $\vect{x}(t)$ evolve according to \eqref{eq:x_system_continuous_C_02}. Denote $\mat{M} = -\mat{I}_{nd} + (\mat{I}_n-\mathcal{L}) \otimes \mat{C}$. Clearly the $j^{th}$ eigenvalue of $\mat{M}$ is equal to $-1 + \lambda_j(\mat{A})$ where $\lambda_i(\mat{A})$ is the $j^{th}$ eigenvalue of $\mat{A} = (\mat{I}_n - \mathcal{L}) \otimes \mat{C}$. The associated eigenvector is $\vect{v}_j$, where $\vect{v}_j$ is the eigenvector of $\mat{A}$ associated with $\lambda_j(\mat{A})$. From \cite[Proposition 7.1.10]{bernstein2009matrixbook}, we conclude that $\lambda_j(\mat{A}) = \mu_i\varphi_k$ where $\mu_i$ and $\varphi_k$ are eigenvalues of $\mat{I}_n-\mathcal{L}$ and $\mat{C}$ respectively, $i\in \mathcal{I}$, $k\in \mathcal{J}$. Then, one can verify that $\vect{v}_j = \vect{u}_i \otimes \vect{w}_k$ is an eigenvector of $\mat{A}$ associated with $\lambda_j(\mat{A})$, where $\vect{u}_i$ and $\vect{w}_k$ are eigenvectors of $\mat{I}_n-\mathcal{L}$ and $\mat{C}$ associated with $\mu_i$ and $\varphi_k$, respectively. According to Assumption~\ref{assm:C_constant}, $\mat{C}$ has a semi-simple eigenvalue at 1 with multiplicity $p\geq 1$; because we need to subsequently distinguish these eigenvalues, we denote them as $\varphi_1, \hdots, \varphi_p$. If $\mathcal{G}$ has a directed spanning tree, then $\mat{I}_n - \mathcal{L}$ has a single eigenvalue at $1$, which we denote as $\mu_1$. Then clearly, $\lambda_j = \mu_1 \varphi_r = 1, r = 1, \hdots, p$ is an eigenvalue of $\mat{A}$ with right eigenvector $\vect{v}_j = \vect{1}_n \otimes \vect{\zeta}_r$. For $\lambda_j = \mu_1 \varphi_k, k = p+1, \hdots, d$, clearly $\lambda_i = \varphi_k$ has real part strictly less than $1$, because Assumption~\ref{assm:C_constant} states that $\mathfrak{Re}(\varphi_k) < 1$. For $\lambda_j = \mu_i \varphi_k$ where $i \in \{2, \hdots, n\}, k \in \mathcal{J}$, if \eqref{cond:stable_C_02} is satisfied then $\lambda_j$ has real part strictly less than $1$. It follows that all eigenvalues of $\mat{M}$ have strictly negative real part, except for $p$ eigenvalues at the origin, with associated right eigenvectors $\vect{v}_j =  \vect{1}_n \otimes \vect{\zeta}_r, r = 1, \hdots, p$. Let $\mat{J}_{\mathcal{L}} = \mat{P}_1^{-1}\mathcal{L}  \mat{P}_1$ and $\mat{J}_{\mat{C}} = \mat{P}_2^{-1} \mat{C} \mat{P}_2$ be the Jordan canonical form of $\mathcal{L}$ and $\mat{C}$, respectively, ordered such that the first Jordan block of $\mat{J}_{\mathcal{L}}$ is associated with the single zero eigenvalue of $\mathcal{L}$ and the first $p$ Jordan blocks of $\mat{J}_{\mat{C}}$ are associated with the $p$ semi-simple unity eigenvalues of $\mat{C}$. With $\mat{P} = \mat{P}_1\otimes \mat{P}_2$, verify that
\begin{equation}
\mat{P}\mat{M}\mat{P}^{-1} = \mat{J} = \begin{bmatrix} \mat 0_{p\times p} & \vect{0}_{p\times (nd-p)} \\ \vect{0}_{(nd-p)\times p} & \mat{\Delta} \end{bmatrix},
\end{equation}
with the $p$ eigenvalues of $\mat{M}$ at the origin being semi-simple and the $nd-p$ nonzero diagonal entries of $\mat{\Delta}$ being the stable eigenvalues of $\mat{M}$. From linear systems theory, one then has that $\vect{x}(t) = e^{\mat{M}t}\vect{x}(0) = \mat{P} e^{\mat{J}t}\mat{P}^{-1}\vect{x}(0)$, which yields $\lim_{t\to\infty} \vect{x}(t) = \sum_{r = 1}^p \vect{p}_r\vect{q}_r^\top\vect{x}(0)$ where $\vect{p}_r$ and $\vect{q}_r^\top$ are right and left eigenvectors of $\mat{M}$ associated with the semi-simple zero eigenvalue, satisfying $\vect{p}_r^\top \vect{q}_r = 1, \forall\,r = 1, \hdots, p$. The above analysis yielded $\vect{p}_r = \vect{1}_n \otimes \vect{\zeta}_r$. One can easily verify that $\vect{q}_r^\top = (\vect{\gamma} \otimes \vect{\xi}_r)^\top$ and thus $\lim_{t\to\infty}\vect{x}(t) = \sum_{r=1}^p(\vect{\gamma}\otimes \vect{\xi}_r)^\top \vect{x}(0) (\vect{1}_n \otimes \vect{\zeta}_r)$. In other words, $\lim_{t\to\infty} \vect{x}_i(t) =  \sum_{r=1}^p\left( \vect{\gamma} \otimes \vect{\xi}_r \right)^\top \vect{x} (0) \vect{\zeta}_r$, which can be rearranged to obtain \eqref{eq:stability_continuous_C_solution}. The sufficiency of \eqref{cond:stable_C_02} has been established. 

It remains for the necessity of \eqref{cond:stable_C_02} to be established. Suppose that \eqref{cond:stable_C_02} is not satisfied. Then there is some $\lambda_j = \mu_i \varphi_k, i \in \{2, \hdots, n\}, k \in \mathcal{J}$ such that the eigenvalue of $\mat{M}$, $-1+\lambda_j$, is in the closed right half-plane. The system is either unstable, or $-1+\lambda_j$ is on the imaginary axis (possibly at the origin). In the latter case either $a)$ there are now at least $p+1$ eigenvalues of $\mat{M}$ at the origin, or $b)$ $\mat{M}$ has a pair of purely imaginary eigenvalues. Regarding $a)$, the system is either unstable (there is a Jordan block of size at least $2\times 2$ in $\mat{J}$ associated with the eigenvalue $0$), or for some $i\neq 1$ and $k \in \mathcal{J}$, there holds $\lambda_j = \mu_i \varphi_k = 0$. Then, $\vect{x}$ converges exponentially fast to a subspace spanned by $\{\vect{v}_1, \hdots, \vect{v}_p, \vect{v}_j\}$ where $\vect{v}_j$ is an eigenvector of $\mat{M}$ associated with eigenvalue $\lambda_j$. Because $i \neq 1$, $\vect{v}_j = \vect{u}_i\otimes \vect{w}_k$ cannot take the form $\vect{1}_n \otimes \vect{w}_k$, for some $\vect{w}_k \in \mathbb{R}^d$, which implies that consensus is not reached for generic initial conditions. Regarding $b)$, denote one of the imaginary eigenvalues as $\lambda_j = \mu_i \varphi_k, i\neq 1$. Then, the system oscillates but not in consensus because, similar to the above arguments, $\vect{v}_j$ associated with the imaginary $\lambda_j$ cannot take the form $\vect{1}_n \otimes \vect{w}_k$. The proof is complete. \hfill$\square$ 

It may be difficult to verify the conditions in Theorem~\ref{thm:stability_continuous_C}  because precise values of eigenvalues of both $\mathcal{L}, \mat{C}$ are needed. We now present two results on sufficient conditions which guarantee consensus \emph{using limited information about the network and the logic structure}. 

\begin{corollary}\label{cor:C_constant_alpha}
	For given $\mat C$ and $\mathcal{G} = (\mathcal{V},\mathcal{E},\mathcal{A})$, suppose that Assumptions~\ref{assm:C_constant} and \ref{assm:graph} are satisfied. Then, there exists a graph $\overline{\mathcal{G}} = \{\mathcal{V}, \mathcal{E}, \overline{\mathcal{A}}\}$ with the same node and edge set as $\mathcal{G}$ but with different edge weights, such that consensus of opinions is achieved using \eqref{eq:xi_update_C}.
\end{corollary}
\emph{\textbf{Proof}:} Let $\mathcal{L}$ be the Laplacian associated with $\mathcal{G}$. Observe that $\mathfrak{Re} \left( ( 1 - \lambda_i(\mathcal{L}) ) \lambda_k ( \mat{C} ) \right)  = d_k - y_i d_k \pm z_i e_k$, 
where, without loss of generality, $\lambda_i(\mathcal{L}) = y_i \pm z_i \jmath$ and $\lambda_k(\mat{C}) = d_k \pm e_k \jmath$ are complex conjugate eigenvalues of $\mathcal{L}$ and $\mat{C}$ respectively, and $z_i, e_k \geq 0$. For $i \in \mathcal{I}\setminus \{1\}$ and $k \in \mathcal{J}$, it follows that $\mathfrak{Re} \left( ( 1 - \lambda_i(\mathcal{L}) ) \lambda_k ( \mat{C} ) \right) < 1 \Leftrightarrow d_k - y_i d_k + z_i e_k < 1$. Define $\overline{\mathcal{A}} = \alpha\mathcal{A}$, where $\alpha > 0$ is a constant scaling every edge weight. Let $\overline{\mathcal{L}}$ be the Laplacian associated with $\overline{\mathcal{G}}$. Since $\mathfrak{Re} \left( ( 1 - \lambda_i(\overline{\mathcal{L}}) ) \lambda_k ( \mat{C} ) \right) = \mathfrak{Re} \left( ( 1 - \alpha\lambda_i(\mathcal{L}) ) \lambda_k ( \mat{C} ) \right)$, it follows that consensus of opinions is achieved on $\overline{\mathcal{G}}$ if and only if $d_k - \alpha(y_i d_k - z_i e_k) < 1\,\forall\,k \in \mathcal{J}$. According to Lemma~\ref{lem:DST_graph}, $z_1= y_1 = 0$, and $y_i > 0$ for all $i \geq 2$. From Assumption~\ref{assm:C_constant}, we have $d_k = 1$ and $e_k = 0$ if $k = 1, \hdots , p$, and $d_k < 1$ otherwise. Thus, there always exists a sufficiently small $\alpha$ satisfying $d_k - \alpha(y_i d_k - z_i e_k) < 1$. 
\hfill$\square$

Next, we present an explicit sufficiency condition which requires limited knowledge of the edge weights of the network, and the logic structure $\mat{C}$.

\begin{corollary}\label{cor:C_constant_stable}
Let $\mat{C}$, which satisfies Assumption~\ref{assm:C_constant}, and $\mathcal{G} = (\mathcal{V},\mathcal{E},\mathcal{A})$ be given. Suppose that $\mathcal{G}$ has a directed spanning tree. Then consensus of opinions is achieved if, for all $k = \{ 1, \hdots, d \}$
\begin{equation}\label{eq:cor_C_constant_if}
\bar{l} < \min \left\{\frac{\vert 1 - \vert \lambda_k\vert \cos(\theta_k)\vert (1+\cos(\theta_k))}{\vert \lambda_k\vert \sin^2(\theta_k)}, 0.5 \right\}
\end{equation}
where $\vert\lambda_k\vert = \vert \lambda_k(\mat{C})\vert$ and $\tan(\theta_k) = e_k/d_k$ with $\lambda_k(\mat{C}) = d_k \pm e_k \jmath$. Here, $\bar{l} = \max_{i\in\mathcal{I}} l_{ii}$ where $l_{ii} = \sum_{j\neq i} a_{ij}$ is the $i^{th}$ diagonal entry of $\mathcal{L}$.
\end{corollary}
\emph{\textbf{Proof}:}  
 From Corollary~\ref{cor:C_constant_alpha}, we recall that the system \eqref{eq:x_system_continuous_C_02} reaches a consensus if and only if 
 \begin{align}\label{eq:cor_iff_cond}
 d_k - y_i d_k + z_i e_k < 1,
 \end{align}
 where $\lambda_i(\mathcal{L}) = y_i \pm z_i \jmath$ and $\lambda_k(\mat{C}) = d_k \pm e_k \jmath$ are any eigenvalue of $\mathcal{L}$ and $\mat{C}$, respectively, except for the case where $\lambda_1(\mathcal{L}) = 0$ and $\lambda_1(\mat{C}) = 1$. According to Assumption~\ref{assm:C_constant}, this means that $d_k < 1$. Recall that $\mathcal{L}$ has nonnegative diagonal entries and nonpositive off-diagonal entries, and moreover each row sums to $0$. Moreover, $\mathcal{L}$ has precisely one eigenvalue at 0 since $\mathcal{G}$ contains a directed spanning tree. Combining these observations with Theorem~\ref{thm:gersgorin}, we conclude that every nonzero eigenvalue of $\mathcal{L}$ is contained in the disc centred at $\bar{l}$, with radius $\bar{l}$. We denote this disc as $D_{\bar{l}}$. The fact that $\bar{l} < 0.5$ implies $y_i < 1$ (from Theorem~\ref{thm:gersgorin}). Thus, $d_k - y_i d_k < 1$ because $d_k < 1$ . This indicates that if $\lambda_k (\mat{C})$ is real, i.e. $e_k = 0$, then \eqref{eq:cor_iff_cond} is satisfied. If all eigenvalues of $\mat{C}$ are real, then $\bar{l} < 0.5$ ensures consensus.

 Consider now $e_k > 0$. Observe that \eqref{eq:cor_iff_cond} is implied by ${z_i}^2 {e_k}^2  < (1 - d_k + y_i d_k)^2$, which is in turn implied by 
 \begin{equation}\label{eq:cor_if_01}
 \bar{z}_i^2 {e_k}^2 < (1 - d_k + y_i d_k)^2
 \end{equation}
 where $\bar{z}_i \geq z_i$ is such that $\beta_i = y_i + \bar{z}_i\jmath$ is on the boundary of $D_{\bar{l}}$. Because $\beta_i$ is on the boundary of $D_{\bar{l}}$, it satisfies $(y_i - \bar{l})^2 + {\bar{z}_i}^2 = {\bar{l}}^2$ which yields ${\bar{z}_i}^2 = - {y_i}^2 + 2 y_i \bar{l}$. Substituting into \eqref{eq:cor_if_01} yields $(- {y_i}^2 + 2 y_i \bar{l}){e_k}^2  < (1 - d_k + y_i d_k)^2 $.
 Expanding and rearranging for $\bar{l}$ yields 
 \begin{align}
 \bar{l} & < \frac{1}{2y_i}\frac{(1-d_k)^2}{{e_k}^2} + \frac{d_k(1-d_k)}{{e_k}^2} + \frac{y_i}{2}\frac{({d_k}^2 + {e_k}^2)}{ {e_k}^2 } 
 \end{align}
 or
 \begin{align}
 \bar{l} & < \frac{ f_k(y_i) }{ \vert \lambda_k\vert^2\sin^2(\theta_k) } \label{eq:cor_if_02}
 \end{align}
 where 
 \begin{align}\label{eq:cor_C_stable_fk}
 f_k(y_i) = \left[ \frac{ (1 - ac)^2 + {y_i}^2c^2 + 2 y_i ac (1 - ac) }{2 y_i}\right]
 \end{align}
 with $a = \cos(\theta_k)$ and $c = \vert \lambda_k\vert$. Recall that $y_i > 0$. Calculations show that $\bar{y}_i= \vert 1-ac\vert/c > 0$ is a unique minimum of $f_k(y_i)$ for $y_i \in (0,\infty)$. Since $f_k(y_i) > 0$ for $y_i \in (0,\infty)$, it follows that \eqref{eq:cor_if_02} is implied by $\bar{l} < f_k(\bar{y}_i)/ \vert \lambda_k\vert^2\sin^2(\theta_k) $, which after some rearranging yields
 \begin{align}\label{eq:cor_if_03}
 \bar{l} < \frac{\vert 1 - \vert \lambda_k\vert \cos(\theta_k)\vert (1+\cos(\theta_k))}{\vert \lambda_k\vert \sin^2(\theta_k)}.
 \end{align}
 The proof is completed by noting that \eqref{eq:cor_if_03} must hold for all $k$ to guarantee that \eqref{eq:cor_C_constant_if} holds. Note that $\vert 1 - \vert \lambda_k \vert \cos(\theta_k) \vert \neq 0$ because $\vert \lambda_k \vert \cos(\theta_k) = d_k < 1$. 

 Consider the scenario where $\mat{C}(\eta)$ varies smoothly as a function of some parameter $\eta \in [a,b]$, and for some $\kappa \in (a,b)$, $\lambda_p(\mat{C}(\kappa))$ has negative real part. Suppose further that $\lambda_p(\mat{C}(\eta))$ is strictly real for $\eta \leq \kappa$, and is complex for $\eta > \kappa$. Then, $\lim_{\eta \to \kappa} \theta_p = \pi$.  Notice that, separately, $\lim_{\theta_p\to\pi} 1+\cos(\theta_p) = 0$ and $\lim_{\theta_p\to\pi} 1+\cos(\theta_p) = 0$. We now show that \eqref{eq:cor_C_constant_if} continues to hold, i.e. is evaluable, as $\theta_p$ approaches $\pi$. Define $g(\theta_p) = \vert \lambda_p \vert \sin^2(\theta_p)$ and $h(\theta_p) = (1 - \vert \lambda_p\vert \cos(\theta_p)) (1+\cos(\theta_p))$. Denote $\lim \theta_p \to \pi^-$ as the limit of $\theta_p$ approaching $\pi$ from the left. Since $h(\theta_p), g(\theta_p)$ are continuous in $\theta_p$, and using L'H\^{o}pital's rule, we obtain via calculations
 \begin{align}
 & \lim_{\theta_p \to { \pi }^- } \frac{h(\theta_p)}{g(\theta_p)}  = \lim_{\theta_p \to { \pi }^- } \frac{h^\prime (\theta_p)}{g^\prime(\theta_p)} = \frac{1+ \vert \lambda_p\vert }{2\vert \lambda_p \vert }.
 \end{align}
 That is, the limit exists. This is consistent with \eqref{eq:cor_C_constant_if} because $(1+\vert\lambda_p\vert)/2\vert\lambda_p\vert > 1/2$ for $\vert\lambda_p\vert > 0$. 
 \hfill$\square$

\begin{remark}
Corollary~\ref{cor:C_constant_alpha} establishes an existence result: there is always a set of edge weights which guarantees consensus. Corollary~\ref{cor:C_constant_alpha} proves this by the scaling of every $a_{ij}$ by a constant $\alpha > 0$, and requires knowledge of $\mat{C}$. In contrast, Corollary~\ref{cor:C_constant_stable} states that we adjust edge weights $a_{ij}$ for individual $i$ only if $l_{ii} = \sum_{j=1}^n a_{ij}$ exceeds the right hand side of \eqref{eq:cor_C_constant_if}, and any such adjustment requires only limited knowledge of the eigenvalues of $\mat{C}$. Moreover, for each individual $i$, the associated $a_{ij}$ need not be scaled by the same constant. While both results need knowledge of $\mathcal{G}$, including the spectral radius of $\mathcal{L}$, we stress that it is only limited knowledge. In the case of Corollary~\ref{cor:C_constant_stable}, limited information concerning $\mat{C}$ is also required.  Additional discussion of the inequality \eqref{eq:cor_C_constant_if}, with simulations, is provided in the following Subsection~\ref{sssec:discuss_v2}.
\end{remark}


\subsection{Convergence in Networks with Stubborn Individuals}
We now study networks with stubborn individuals, i.e. $\exists\,i \in\mathcal{I}: b_i > 0$. We first give a standard result for the convergence of an exponentially stable linear system with a constant input.
	\begin{lemma}\label{lem:input_limit}
		Consider the linear system $\dot{\vect x}(t) = -\mat F \vect x(t) + \vect u$, where $-\mat F$ is Hurwitz, and $\vect u$ is a constant vector. Then, $\lim_{t\to\infty} \vect x(t) = \mat F^{-1}\vect u$ exponentially fast.
	\end{lemma}
\emph{\textbf{Proof}:} 
 Under the lemma hypotheses, $-\m F$ is Hurwitz. Linear systems theory states that the solution of \eqref{eq:model_stubborn} is given by
 \begin{align} \label{eq:solution_stubborn}
     \m{x}(t)   = \Big({e}^{-\mat F t}\m{x}(0) + \int_{0}^t {e}^{-\mat F (t - \tau)} \mathrm{d}
     \tau \m{u} \Big) .
 \end{align}
 Taking $t\to\infty$ on both sides of Eq.~\eqref{eq:solution_stubborn} yields
 \begin{align}
     \lim_{t\to\infty} \m{x}(t)  & =  \lim_{t\to\infty} {e}^{-\m F t}\Big(\int_{0}^t {e}^{\m F\tau} \mathrm{d}\tau \Big)\m{u}  \label{eq:stub_3}
 \end{align}
 We claim that $\int_{0}^t {e}^{\m F\tau} \mathrm{d}\tau = \m F^{-1}(e^{\m F t} - \m I_{nd})$. To see this, observe that ${e}^{\m F\tau} \triangleq \sum_{k=0}^{\infty} \frac{(\m F\tau)^k}{k!} $ implies $\int_{0}^t {e}^{\m F\tau} \mathrm{d}\tau = t\big[ \m I_{nd} + \sum_{k=2}^{\infty} \frac{(\m Ft)^{k-1}}{k!} \big]$. One obtains  $\big(\int_{0}^t {e}^{\m F\tau} \mathrm{d}\tau\big)\m F + \m I_{nd} = e^{\m F t}$, and the invertibility of $\m F$ yields $\int_{0}^t {e}^{\m F\tau} \mathrm{d}\tau = \m F^{-1}(e^{\m F t} - \m I_{nd})$. Thus, Eq.~\eqref{eq:stub_3} becomes
 \begin{align}
     \lim_{t\to\infty} \m{x}(t)  & = \lim_{t\to\infty} ({e}^{-\m F t} \m F^{-1}e^{\m F t} - {e}^{-\m F t}\m F^{-1})\m{u} \nonumber \\
     & = \m F^{-1}\m{u} .
 \end{align}\hfill $\qed$ 

	Consider the system \eqref{eq:compact_network_stubborn_02}. By establishing conditions for which the matrix 
	\begin{align}\label{eq:compact_network_stubborn_unforced}
	\bar{\mat M} = -\left[\mat{I}_{nd} + \big((\mathcal{L}- \mat{I}_n)\otimes \mat{C}\big) + \mat{B} \otimes \mat{I}_d \right]
	\end{align}
	is Hurwitz, by treating $(\mat B \otimes \mat I_d)\vect x(0)$ as a constant input, and replacing $\mat F$ with $\bar{\mat M}$, Lemma~\ref{lem:input_limit} allows us to establish conditions for which
	\begin{align}\label{eq:x_final_stub}
	\lim_{t\to\infty} \vect{x}(t) & = \left[\mat{I}_{nd} + (\mathcal{L}- \mat{I}_n)\otimes \mat{C} + \mat{B} \otimes \mat{I}_d \right]^{-1} \nonumber \\
	& \quad \quad \times (\mat{B}\otimes \mat{I}_d)\vect{x}(0).
	\end{align}
Note that if $\vect{x}_i(0) = \vect{x}_j(0),\forall i,j\in \mathcal{I}$, i.e. all individuals are initially at consensus, then clearly $\dot{\vect x} = (\mat{I}_n\otimes (\mat{C}-\mat{I}_d))\vect{x}$ and $\lim_{t\to\infty}\vect{x}_i(t) = \sum_{k=1}^p \vect{\xi}_k^\top\vect{x}_i(0)\vect{\zeta}_k$ for all $i\in\mathcal{I}$, where $\vect{\xi}_k^\top$ and $\vect{\zeta}_k$ were given in Assumption~\ref{assm:C_constant}. When the initial conditions are not equal, the opinions converge to \eqref{eq:x_final_stub}, which in general corresponds to a persistent disagreement of opinions. In what follows, we present results for individuals who (i) are slightly stubborn, (ii) have approximately the same stubbornness, and (iii) are extremely stubborn. 

\begin{theorem}\label{thm:stability_small_stub}
For given $\mat C$ and $\mathcal{G} = (\mathcal{V},\mathcal{E},\mathcal{A})$, suppose that Assumptions~\ref{assm:C_constant} and \ref{assm:graph} are satisfied. Suppose further that \eqref{cond:stable_C_02} is satisfied. Then, the opinion dynamics system \eqref{eq:compact_network_stubborn_02} with stubborn individuals converges to \eqref{eq:x_final_stub} if
	\begin{enumerate}
		\item Parameter $b_i \geq 0$ is sufficiently small, for all $i \in \mathcal{I}$, and $\exists j \in \{1, \hdots, r\}: b_j > 0$.
		\item For some $\alpha > 0$, $b_i = \alpha + \epsilon_i$ for some sufficiently small $\epsilon_i\in \mathbb{R}, \forall\,i\in \mathcal{I}$. 
	\end{enumerate} 
\end{theorem}
\emph{\textbf{Proof}:} 
\textit{Item 1:} In the proof of Theorem~\ref{thm:stability_continuous_C}, we established that if consensus is reached for the system \eqref{eq:x_system_continuous_C_02}, then $\mat{M} = -\mat{I}_{nd} + \big((\mat{I}_n-\mathcal{L})\otimes \mat{C}\big)$ has a single eigenvalue at zero. We denoted this as $\lambda_1(\mat{M}) = 0$, and showed in that same proof that $\lambda_1(\mat{M}) = 0$ has an associated left eigenvector $\vect{u}_1 = \vect{1}_n\otimes \vect{\zeta}$ and right eigenvector $\vect{v}_1^\top = \vect{\gamma}^\top \otimes \vect{\xi}^\top$ where $\vect{\xi}^\top$ and $\vect{\zeta}$ are given in Assumption~\ref{assm:C_constant}, and $\vect{\gamma}^\top$ is detailed below Lemma~\ref{lem:DST_graph}.

We now establish the exponential stability of the system \eqref{eq:compact_network_stubborn_unforced}. Define $\vect{b} = [b_1, \hdots, b_n]^\top$. Next, by defining
\begin{equation}\label{eq:thm_stubborn_Z_def}
\mat{Z}(\vect{b}) = -\mat{I}_{nd} + \big((\mat{I}_n - \mathcal{L})\otimes \mat{C}\big) - \mat{B}(\vect{b}) \otimes \mat{I}_d 
\end{equation}
observe that the system \eqref{eq:compact_network_stubborn_unforced} is equivalent to the system $\dot{\vect{x}} = \mat{Z}(\vect{b})\vect{x}$. Given a simple eigenvalue $\lambda_k(\mat{Z}(\vect b))$ of $\mat{Z}(\vect b)$, with associated left and right eigenvectors $\vect{u}_k(\vect b)$ and $\vect{v}_k(\vect b)$ satisfying $\vect{u}_k(\vect b)^\top\vect{v}_k(\vect b) = 1$,
observe that, with $b_i \geq 0$ and $\mathbf{e}_i \in \mathbb{R}^n$ defined at the start of Section~\ref{section:background_problem}, there holds 
\begin{equation}\label{eq:thm_stubborn_gradient_NSD}
\frac{\partial}{\partial b_i } \left[\mat{Z}(\vect{b})\right] = - \mathbf{e}_i \mathbf{e}_i^\top \otimes \mat{I}_d \leq 0,
\end{equation}
where $\mathbf{e}_i \in \mathbb{R}^n$ is the canonical unit vector in the $i^{th}$ dimension.

Consider now any $k \in \{1, \hdots, nd\}$, and observe that $\mat{Z}(\vect{b})\vect{v}_k(\vect{b}) = \lambda_k(\mat{Z}(\vect{b})) \vect{v}_k(\vect{b})$ where $\lambda_k(\mat{Z}(\vect{b}))$ is the $k^{th}$ eigenvalue of $\mat{Z}(\vect{b})$ with associated left and right eigenvectors $\vect{u}_k(\vect{b})$ and $\vect{v}_k(\vect{b})$, respectively. For convenience, we drop the argument $\vect{b}$ when there is risk of confusion, but we stress that the matrix $\mat{Z}(\vect{b})$ and its eigenvalues and eigenvectors are functions of $\vect{b}$. We will now investigate the derivatives of $\mat{Z}$, $\lambda_k(\mat{Z})$, $\vect{u}_k$ and $\vect{v}_k$ with respect to $b_i, i \in \{1, \hdots, nd\}$. Observe further that, for any $i\in \mathcal{I}$, we have $\frac{\partial}{\partial b_i }\left[\mat{Z}\vect{v}_k\right] = \frac{\partial}{\partial b_i } \left[ \lambda_k(\mat{Z}) \vect{v}_k \right]$ which is equivalent to 
\begin{align*}\label{eq:thm_stubborn_derivative_01}
\frac{\partial}{\partial b_i } & \left[\mat{Z} \right]\vect{v}_k \!+\! \mat{Z} \frac{\partial}{\partial b_i } \left[ \vect{v}_k \right] \!=\! \frac{\partial}{\partial b_i } \left[ \lambda_k(\mat{Z}) \right]\vect{v}_k + \lambda_k(\mat{Z}) \frac{\partial}{\partial b_i } \left[ \vect{v}_k \right]
\end{align*}
Assume without loss of generality that $\vect{u}_k, \vect{v}_k$ are normalised such that $\vect{u}_k^\top\vect{v}_k = 1$. Premultiplying both sides of the above equation by $\vect{u}_k^\top$ yields
\begin{align}
\vect{u}_k^\top& \frac{\partial}{\partial b_i } \left[\mat{Z} \right]\vect{v}_k + \vect{u}_k^\top\mat{Z}\frac{\partial}{\partial b_i }\left[ \vect{v}_k \right] \nonumber \\ 
& = \vect{u}_k^\top\frac{\partial}{\partial b_i }\left[ \lambda_k(\mat{Z}) \right]\vect{v}_k + \lambda_k(\mat{Z})\vect{u}_k^\top\frac{\partial}{\partial b_i }\left[ \vect{v}_k \right].
\end{align}
By recalling that $\vect{u}_k^\top\mat{Z} = \lambda_k(\mat{Z}) \vect{u}_k(\vect{b})^\top$, we see that the second term on the left hand side cancels the second term on the right hand side of the above. Additionally, $\vect{u}_k^\top\frac{\partial}{\partial b_i } \left[ \lambda_k(\mat{Z}) \right] \vect{v}_k = \frac{\partial}{\partial b_i } \left[ \lambda_k(\mat{Z}) \right]$ because we assumed the normalisation $\vect{u}_k^\top\vect{v}_k = 1$. This yields \begin{align*}\label{eq:thm_stubborn_derivative_02}
\frac{\partial}{\partial b_i } \left[ \lambda_k(\mat{Z}) \right] & =  \vect{u}_k^\top \frac{\partial}{\partial b_i } \left[\mat{Z} \right]\vect{v}_k .
\end{align*}
In the proof of Theorem~\ref{thm:stability_continuous_C}, we showed that $\lambda_1(\mat{Z}(\vect{0}_n)) = 1$ is a simple eigenvalue and has associated left and right eigenvectors $\vect{u}_1 = \vect{\gamma}^\top \otimes \vect{\xi}^\top$ and $\vect{v}_1 = \vect{1}_n\otimes \vect{\zeta}$. From Lemma~\ref{lem:DST_graph} and the arguments below it, we recall that $\vect{\gamma}^\top = [\gamma_1, \hdots, \gamma_n]$ has nonnegative elements, with $\gamma_i > 0, i \in \{1,\hdots, r\}$. It then follows that 
\begin{align}
\frac{\partial}{\partial b_i } \left[ \lambda_1(\mat{Z}(\vect{0}_n)) \right] & = - (\vect{\gamma}^\top\otimes\vect{\xi}^\top) (\mathbf{e}_i\mathbf{e}_i^\top \otimes \mat{I}_d)(\vect{1}_n \otimes \vect{\zeta}) \nonumber \\
& = - \gamma_i \leq 0.
\end{align}
because $\frac{\partial}{\partial b_i } \left[\mat{Z}(\vect{b})\right] = - \mathbf{e}_i \mathbf{e}_i^\top \otimes \mat{I}_d.$ Thus, for sufficiently small $b_i \geq 0, i\in\mathcal{I}$, the gradient $\frac{\partial}{\partial b_i } \left[ \lambda_1(\mat{Z}(\vect{0}_n)) \right]$ is nonpositive. Moreover, because we assumed that $\exists\,j \in \{1, \hdots, r\}: b_j > 0 \Rightarrow \frac{\partial}{\partial b_j }\left[ \lambda_1(\mat{Z}(\vect{0}_n)) \right] = -\gamma_j < 0$, the eigenvalue $\lambda_1(\mat{Z}(\vect{0})) = 0$ moves into the open left half-plane as $b_i$ increases from $0$. In other words, for sufficiently small $b_j$ and $b_i$, $\lambda_1(\mat{Z}(\vect{b}))$ becomes a stable eigenvalue. The other $nd-1$ eigenvalues are continuous functions of $\vect{b}$ and thus will remain in the open left-half plane for small $b_i$ (this is because the eigenvalues are already in the open left half-plane by virtue of the fact that the nonstubborn system \eqref{eq:x_system_continuous_C_02} is assumed to reach a consensus). We conclude also that $\mat{Z}(\vect{b})$ is nonsingular, and on recalling the definition of $\mat{Z}(\vect{b})$ given in \eqref{eq:thm_stubborn_Z_def}, completes the proof for \textit{Item 1}.

\textit{Item 2:} Suppose first that $\epsilon_i = 0\,\forall\,i$. If $b_i = \alpha\forall\,i$, then \eqref{eq:compact_network_stubborn_unforced} yields
\begin{align}\label{eq:compact_network_stubborn_unforced_alpha}
\dot{\vect{x}} & = \big[-(1+\alpha)\mat{I}_{nd} - (\mathcal{L} - \mat{I}_n)\otimes \mat{C} \big]\vect{x} 
\end{align}
which implies that the eigenvalues of $\mat{Z} = -(1+\alpha)\mat{I}_{nd} - (\mathcal{L}- \mat{I}_n)\otimes \mat{C}$ are the eigenvalues of $-\mat{I}_{nd} - (\mathcal{L}- \mat{I}_n)\otimes \mat{C}$ (which are all in the open left half-plane except for one at the origin) shifted along the real axis by $-\alpha < 0$, which ensures the exponential stability of \eqref{eq:compact_network_stubborn_unforced_alpha} and the final opinions converge to \eqref{eq:x_final_stub}. 

Next, we consider $\epsilon_i \neq 0$ for some $i$. From the fact that the eigenvalues of $\mat{Z}$ are continuous functions of $\epsilon_i$, we conclude that for sufficiently small $\epsilon_i$, all eigenvalues of $\bar{\mat{Z}} = -(1+\alpha)\mat{I}_{nd} - (\mathcal{L}- \mat{I}_n)\otimes \mat{C} - \sum_{i=1}^{n} \epsilon_i \mathbf{e}_i \mathbf{e}_i^\top \otimes \mat{I}_d$ will remain in the open left-half plane. In other words, for minor perturbations induced by $\epsilon_i$, $\bar{\mat{Z}}$ remains stable.
 \hfill$\square$
 
 \begin{lemma}\label{lem:stability_large_stub}
 	For given $\mat C$ and $\mathcal{G} = (\mathcal{V},\mathcal{E},\mathcal{A})$, suppose that Assumptions~\ref{assm:C_constant} and \ref{assm:graph} are satisfied. Then, the opinion dynamics system \eqref{eq:compact_network_stubborn_02} with stubborn individuals converges to \eqref{eq:x_final_stub} if $b_i > 0$ is sufficiently large, for all $i\in \mathcal{I}$.
 \end{lemma}
\emph{\textbf{Proof}:} 
The proof is an application of Ger\v{s}gorin's Circle Theorem. Examination of \eqref{eq:compact_network_stubborn_unforced} shows that $b_i > 0$ does not change the size of the associated $(i-1)d+1, (i-1)d, \hdots, (i-1)d+d$ Ger\v{s}gorin disks, but does shift the centre of the disc further along the negative real axis towards $-\infty$. If every $b_i$ is sufficiently large, then every Ger\v{s}gorin disk will be strictly inside the open left half-plane, indicating that the system \eqref{eq:compact_network_stubborn_unforced} is exponentially stable about the origin $\vect{x} = \vect{0}_{nd}$. This, along with the arguments preceding \eqref{eq:x_final_stub}, completes the proof. \hfill$\square$

\begin{remark}
	Lemma~\ref{lem:stability_large_stub} does not require the system \eqref{eq:x_system_continuous_C_02} to reach a consensus: high stubbornness in all individuals ensures the opinion dynamics process is convergent even when the topic couplings are complex. However, this may be at the expense of reaching a consensus. Future work may study an adaptive $b_i(t)$, capturing individuals who increase their stubbornness if the opinion evolution process is becoming unstable (an undesirable scenario). 
\end{remark}

\subsection{Convergence for a Class of $\mat{C}$ Matrices}\label{ssec:class_C}
In many opinion dynamics problems, it is desirable to scale the opinions to be in some predefined interval $[-a, a]$, for some scalar $a > 0$, and one \emph{desirable property of an opinion dynamics model} is that opinions starting in $[-a, a]$ remain inside $[-a, a]$ for all time \cite{parsegov2017_multiissue}. Supposing that the $k^{th}$ topic represents the discussion of attitudes towards a statement, e.g. ``recreational marijuana should be legal'', one might scale the opinions so that $x_i^k = a$ represents maximal support for the statement, $x_i^k = 0$ represents a neutral stance, while $x_i^k = -a$ represents maximal rejection of the statement. Now, we explore one set of sufficient requirements on $\mathcal{G}$ and the $\mat{C}$ which ensures that \eqref{eq:xi_update_C_stub} has this desirable property.

\begin{assumption}\label{assm:L_C_invariant}
	The $i^{th}$ diagonal entry of the Laplacian matrix $\mathcal{L}$, associated with $\mathcal{G} = (\mathcal{V}, \mathcal{E}, \mathcal{A})$, satisfies $l_{ii} \leq 1,\forall i$. The $k^{th}$ diagonal of the logic matrix $\mat{C}$ satisfies $c_{kk} > 0$ for all $k$, and $\Vert \mat{C}\Vert_{\infty} = 1$.
\end{assumption}
Note that the constraint on the Laplacian entries $l_{ii}$ can always be satisfied by scaling the adjacency matrix weights. We show an invariant set property for \eqref{eq:compact_network_stubborn_02}, and then consider networks without stubborn individuals, and then with stubborn individuals, i.e. \eqref{eq:compact_network_stubborn_02} with $\mat B = \mat 0_{n\times n}$ and then with $\mat B \neq \mat 0_{n\times n}$, respectively.


\begin{lemma}
Suppose that Assumptions~\ref{assm:C_constant}, \ref{assm:graph}, and \ref{assm:L_C_invariant} hold for given $\mat{C}$ and $\mathcal{G}=(\mathcal{V}, \mathcal{E}, \mathcal{A})$. Suppose further that each individual's opinion evolves according to \eqref{eq:xi_update_C_stub}. If $\vect{x}(0) \in \mathcal{R} \triangleq \{\vect{x} : x_i^k \in [-a, a],\forall i \in \mathcal{I}, \forall k \in \mathcal{J}\}$ for some arbitrary but fixed $a \in \mathbb{R}_{+}$, then $\vect{x}(t) \in \mathcal{R}$ for all $t\geq 0$. 
\end{lemma}
\emph{\textbf{Proof}:} 
As above \eqref{eq:xi_update_C_stub}, we define the $k^{th}$ opinion of individual $i$ as $x_i^k(t)$. To prove the lemma statement, we need only prove that, for all $k\in \mathcal{J}$ and $i\in \mathcal{I}$, there holds
\begin{align}
\dot{x}_i^k(t) & \leq 0,\quad \text{if} \quad x_i^k(t) = a \label{eq:invariant_upper} \\
\dot{x}_i^k(t) & \geq 0,\quad \text{if} \quad x_i^k(t) = -a \label{eq:invariant_lower}
\end{align}
for $\vect{x}(t) \in \mathcal{R}$. Denote the $k^{th}$ row of $\mat{C}$ as $\vect{c}_k^\top$. Dropping the $t$ argument for clarity, we obtain from \eqref{eq:xi_update_C_stub}:
\begin{align}
\dot{x}_i^k & = \sum_{j\in\mathcal{N}_i} a_{ij} \vect{c}_k^\top (\vect{x}_j - \vect{x}_i) + \vect{c}_k^\top\vect{x}_i - x_i^k + b_i(x_i^k(0)-x_i^k) \nonumber \\
& = \sum_{j\in\mathcal{N}_i} a_{ij} \sum_{l=1}^d c_{kl} (x_j^l - x_i^l) + \sum_{l=1}^d c_{kl} x_i^l  \nonumber \\
& \quad \quad - x_i^k + b_i(x_i^k(0)-x_i^k) \nonumber \\
& = \sum_{j\in\mathcal{N}_i} a_{ij} \sum_{l=1}^d c_{kl} x_j^l + (1 - l_{ii})\sum_{l=1}^d c_{kl} x_i^l \nonumber \\
& \quad \quad  - x_i^k + b_i(x_i^k(0)-x_i^k) \label{eq:dot_xi}
\end{align}
for any $k \in \mathcal{J}$ and $i\in \mathcal{I}$, with the last equality obtained by noting that $\sum_{j\in\mathcal{N}_i} a_{ij} = l_{ii}$. With $x_i^k = a$ and $\vect{x}(t) \in \mathcal{R}$, we obtain $b_i(x_i^k(0)-x_i^k) \leq 0$. It follows from \eqref{eq:dot_xi} that \eqref{eq:invariant_upper} holds if
\begin{align}
a(1-l_{ii})c_{kk} - a & + (1-l_{ii})  \sum_{l=1, l \neq k}^d c_{kl} x_i^l \nonumber \\
& + \sum_{j\in\mathcal{N}_i} a_{ij} \sum_{l=1}^d c_{kl} x_j^l \leq 0 .\label{eq:dot_xi_ineq_01}
\end{align}
Observe that \eqref{eq:dot_xi_ineq_01} is implied by 
\begin{align}
(1-l_{ii})c_{kk} - 1 + \vert 1- l_{ii} \vert \sum_{l=1,l\neq k}^d \vert c_{kl} \vert + l_{ii} \sum_{l=1}^d \vert c_{kl} \vert \leq 0 \label{eq:dot_xi_ineq_02}
\end{align}
because $\vert x_j^l \vert \leq a,\forall\, j\in \mathcal{I}, l \in \mathcal{J}$ (including $x_i^l$) and $\sum_{j\in\mathcal{N}_i} a_{ij} = l_{ii}$. Since $l_{ii} \leq 1$ and $c_{kk} + \sum_{l=1,l\neq k}^d \vert c_{kl} \vert  = \sum_{l=1}^d \vert c_{kl} \vert \leq 1$ under Assumption~\ref{assm:L_C_invariant}, \eqref{eq:dot_xi_ineq_02} evaluates to be
\begin{align}
(1-l_{ii})\Big( c_{kk} +\! \sum_{l=1,l\neq k}^d \vert c_{kl}\vert \Big) \!+ l_{ii} \sum_{l=1}^d \vert c_{kl}\vert \!-\! 1 & \leq 0 . \label{eq:dot_xi_ineq_03}
\end{align}
It follows that for all $t\geq 0$, for any $k \in \mathcal{J}$ and $i \in \mathcal{I}$, the inequality in \eqref{eq:dot_xi_ineq_03} holds. Because \eqref{eq:dot_xi_ineq_03} holds, then \eqref{eq:dot_xi_ineq_01} holds, and thus \eqref{eq:invariant_upper} holds. One can use the same approach to obtain a similar proof for \eqref{eq:invariant_lower}. The proof is complete.\hfill$\square$


\begin{theorem}[No Stubborn Individuals]\label{thm:consensus_assm2}
Suppose that Assumptions~\ref{assm:C_constant} and \ref{assm:L_C_invariant} hold for $\mat{C}$ and $\mathcal{G} = (\mathcal{V}, \mathcal{E}, \mathcal{A})$. Then, with each individual's opinions evolving according to \eqref{eq:xi_update_C}, the network of individuals globally exponentially reaches a consensus on all topics, with final opinions given in \eqref{eq:stability_continuous_C_solution}, if and only if $\mathcal{G}$ has a directed spanning tree. 
\end{theorem}

\emph{\textbf{Proof}:} 
First, we note that the proof of Theorem~\ref{thm:stability_continuous_C} established that \eqref{cond:stable_C_02} holds only if $\mathcal{G}$ has a directed spanning tree. This establishes the necessity of the directed spanning tree. Before proving sufficiency, we first derive some properties of the eigenvalues of $\mat{M} = -\mat{I}_{nd} + (\mat{I}_n-\mathcal{L}) \otimes \mat{C}$. Consider a given $l\in \{1, \hdots, nd\}$. The $l^{th}$ diagonal entry of $\mat{M}$ is $m_{ll} = -1 + (1-\sum_{j\in\mathcal{N}_i} a_{ij})c_{kk}$ for some $i\in\mathcal{I}$ and $k\in \mathcal{J}$. The off-diagonal entries of the $l^{th}$ row, $m_{lj}$, are given by $(1-l_{ii})c_{kp}$ for all $q\in\mathcal{I}, p\in \mathcal{J}, p \neq k$, and $a_{iq}c_{kp}$ for all $q\in\mathcal{I}, p\in \mathcal{J}$. From Assumption~\ref{assm:L_C_invariant}, we have  $ 0 < c_{kk} \leq 1$ and $\sum_{j\in\mathcal{N}_i} a_{ij} = l_{ii} \leq 1\,\forall\,i \in \mathcal{I} \Rightarrow 0 \leq 1 - \sum_{j\in\mathcal{N}_i} a_{ij} \leq 1$. It follows that $m_{ll} \leq 0$ for all $l \in \{1, \hdots, nd\}$. As in Theorem~\ref{thm:gersgorin}, define $R_l(\mat{M}) = \sum_{j=1,j\neq l}^{nd} \vert m_{lj} \vert$, i.e. the sum of the absolute values of the off-diagonal entries of the $l^{th}$ row of $\mat{M}$. Observe that
\begin{align}
R_l(\mat{M}) & = \vert 1-l_{ii}\vert \sum^d_{p=1,p\neq k} \vert c_{kp}\vert  + \sum_{j\in\mathcal{N}_i} a_{ij} \hat c_k \nonumber \\
& = (1-l_{ii}) \sum^d_{p=1,p\neq k} \vert c_{kp}\vert  + l_{ii} \hat c_k,
\end{align}
where $\hat c_k = \sum^d_{p=1} \vert c_{kp} \vert$ is the sum of the absolute values of the $k^{th}$ row of $\mat{C}$. Note that $0 \leq 1-l_{ii} \leq 1$. Thus,
\begin{align}
& m_{ll} +  R_l(\mat{M}) = \nonumber \\
& \quad -1 + (1-l_{ii})c_{kk} + (1-l_{ii}) \sum^d_{p=1,p\neq k} \vert c_{kp}\vert  + l_{ii} \hat c_k \nonumber \\
& = -1 + (1-l_{ii})\Big( \sum^d_{p=1, p\neq k} \vert c_{kp} \vert + c_{kk}\Big) + l_{ii} \hat c_k .
\end{align}
From Assumption~\ref{assm:L_C_invariant}, we have $\Vert \mat{C} \Vert_{\infty} =1$ and $c_{kk} > 0$, which implies that $\sum^d_{p=1, p\neq k} \vert c_{kp} \vert + c_{kk} = \hat c_k \leq 1$. It follows that 
\begin{equation}
m_{ll} + R_l(\mat{M}) =  - 1 + \hat c_k\leq 0.
\end{equation}
This implies that $m_{ll} \leq - R_l(\mat{M})$, and that this holds for all $l \in \{1, \hdots, nd\}$. Thus, the Ger\v{s}gorin discs of $\mat{M}$ are all in the left half-plane. Specifically, the discs are either in the open left half-plane ($m_{ll} < - R_l(\mat{M})$) or touch the imaginary axis at the origin but do not enclose it ($m_{ll} = - R_l(\mat{M})$, with this including the possibility that $m_{ll} = 0$). This implies that the eigenvalues of $\mat{M}$ either have strictly negative real part, or are equal to zero. Define $\mat{A} = (\mat{I}_n-\mathcal{L}) \otimes \mat{C}$, with eigenvalue $\lambda_i = (1-\mu_k)\varphi_l$, where $\mu_k$ and $\varphi_l$ are eigenvalues of $(\mat{I}_n-\mathcal{L})$ and $\mat{C}$, respectively. 

The analysis in the proof of Theorem~\ref{thm:stability_continuous_C} showed that if $\mathcal{G}$ has a directed spanning tree, then satisfying \eqref{cond:stable_C_02} ensured a consensus was reached. Because the eigenvalues of $\mat{M}$ either have strictly negative real part, or are equal to zero, we need only to show that $\lambda_i(\mat{A}) = \mu_k \varphi_l \neq 1$ for all $k \in \{2, \hdots, n\}$ and $l \in \mathcal{J}$ to satisfy \eqref{cond:stable_C_02}.
Because $\mathcal{L}$ has a simple zero eigenvalue with multiplicity $p\geq 1$ and all other eigenvalues have positive real part, it follows that $\mathfrak{Re}(\mu_k) < 1$ for $k > p$. This implies that $\lambda_i(\mat{A}) = \mu_k\varphi_j \neq 1$, for all $k \neq 1, j = 1, \hdots, p$ according to Assumption~\ref{assm:C_constant}. Consider now $k \in \{2, \hdots, n\}$ and $l \in \{p+1, \hdots, d\}$. Because $l_{ii} \leq 1$, we conclude using Theorem~\ref{thm:gersgorin} that $\vert \mu_k \vert \leq 1$. Because $c_{ll} > 0,\forall\,l$ and $\Vert \mat{C} \Vert_{\infty} = 1$, the $l^{th}$ Ger\v{s}gorin disc of $\mat{C}$ is situated at $c_{ll}$ with radius $1 - c_{ll}$. It follows that $\vert \varphi_l \vert < 1$ for $l = \{p+1, \hdots, d\}$. Thus $\vert \lambda_i\vert = \vert\mu_k \varphi_l \vert \leq \vert \mu_k \vert \vert \varphi_l\vert < 1$ for all $k \in \{2, \hdots, n\}$ and $l\in \{ p+1, \hdots, d\}$. In other words, \eqref{cond:stable_C_02} of Theorem~\ref{thm:stability_continuous_C} is satisfied. The final consensus value is computed as in the proof of Theorem~\ref{thm:stability_continuous_C}. \hfill$\square$


\begin{theorem}[Stubborn Individuals]\label{thm:stub_assm2}
Suppose that Assumption~\ref{assm:C_constant}, \ref{assm:graph}, and \ref{assm:L_C_invariant} hold for given $\mat{C}$ and $\mathcal{G}=(\mathcal{V}, \mathcal{E}, \mathcal{A})$. If each individual's opinions evolve according to \eqref{eq:xi_update_C_stub}, $b_i\geq 0,\forall\,i\in\mathcal{I}$, and $\exists j\in \{1, \hdots, r\} : b_j > 0$, then, the opinion dynamics system \eqref{eq:compact_network_stubborn_02} converges to \eqref{eq:x_final_stub} exponentially fast.
\end{theorem}
\emph{\textbf{Proof}:} Notice that $\bar{\mat{M}}$ in \eqref{eq:compact_network_stubborn_unforced} can be expressed as $\bar{\mat M} = \mat{M} - \mat{B}\otimes \mat{I}_d$, where $\mat{M} = -\mat{I}_{nd} + (\mat{I}_n-\mathcal{L}) \otimes \mat{C}$ was defined in the proof of Theorem~\ref{thm:stability_continuous_C}. We showed in that proof that the Ger\v{s}gorin discs of $\mat{M}$ were in the closed left half-plane, and the discs could at most touch the origin, but not enclose it. This implied that $\mat{M}$ has eigenvalues that either have strictly negative real part, or are at the origin. If $b_i\geq 0,\forall\,i\in\mathcal{I}$ then $-\mat{B}\otimes \mat{I}_d$ is a diagonal matrix with nonpositive diagonal entries. It follows that, for $i\in\mathcal{I}$, the $(i-1)d+1, \hdots, (i-1)d+d$ Ger\v{s}gorin discs of $\bar{\mat M}$ are the $(i-1)d+1, \hdots, (i-1)d+d$ Ger\v{s}gorin discs of $\mat M$, with the same radius, but shifted along the real axis to the left by $b_i \geq 0$. 
Thus, by proving that $\bar{\mat{M}}$ is invertible (as we shall now do) it follows that all eigenvalues of $\bar{\mat{M}}$ have negative real part.

To establish a contradiction, suppose that $\bar{\mat{M}} = -\mat{I}_{nd} + (\mat{I}_n-\mathcal{L}) \otimes \mat{C} - \mat{B}\otimes \mat{I}_d$ is not invertible. Then there exists a nonzero vector $\vect{x} \in \mathbb{R}^{nd}$ such that $\bar{\mat M} \vect x = \vect 0_{nd}$. This implies that $\left((\mat{B}+\mat{I}_n)\otimes \mat{I}_d\right)\vect{x} = \left( (\mat{I}_n - \mathcal{L}) \otimes \mat{C} \right) \vect{x} $, or
\begin{align}
\vect{x} & = \left( ((\mat{B}+\mat{I}_n)^{-1}(\mat{I}_n - \mathcal{L}))\otimes \mat{C} \right) \vect{x} \label{eq:eig_1_N}
\end{align}
with $\mat{B}+\mat{I}_n$ always invertible because $b_i \geq 0$ for all $i\in\mathcal{I}$. Obviously, \eqref{eq:eig_1_N} holds if and only if the matrix $\mat{N} = ((\mat{B}+\mat{I}_n)^{-1}(\mat{I}_n - \mathcal{L}))\otimes \mat{C}$ has an eigenvalue at 1. Notice that the $(i-1)d+1, \hdots, (i-1)d+d$ rows of $(\mat{B}+\mat{I}_n)^{-1}(\mat{I}_n-\mathcal{L})$ are equal to the rows of $\mat{I}_n-\mathcal{L}$ scaled by $(b_i+1)^{-1} \leq 1$. Using Theorem~\ref{thm:gersgorin}, we first conclude the eigenvalues of $\mat{I}_n-\mathcal{L}$ are in the closed unit circle, and then conclude that the eigenvalues of $(\mat{B}+\mat{I}_n)^{-1}(\mat{I}_n-\mathcal{L})$ are in the closed unit circle.

We prove by contradiction that $\mat{N}$ does not have an eigenvalue at $1$. Denote the eigenvalues of $\mat{N}$, $(\mat{B}+\mat{I}_n)^{-1}(\mat{I}_n-\mathcal{L})$, and $\mat{C}$ as $\psi_i$, $\bar{\mu}_k$, and $\varphi_l$, and \cite[Proposition 7.1.10]{bernstein2009matrixbook} indicates that $\psi_i = \bar{\mu}_k\varphi_l$, for $k\in\mathcal{I}$ and $l \in \mathcal{J}$. Note also that, from Assumption~\ref{assm:C_constant}, $\varphi_r = 1, r = 1, \hdots, p$ is a semi-simple eigenvalue of $\mat{C}$. Using the same arguments as in the last paragraph of the proof of Theorem~\ref{thm:consensus_assm2}, one can establish that under Assumption~\ref{assm:L_C_invariant}, $\psi_i = \bar{\mu}_k\varphi_l \neq 1$ for $k\in\mathcal{I}$ and $l\in \{p+1, \hdots, d\}$. Thus, $\mat{N}$ has an eigenvalue at $1$ if and only if $\psi_i = \bar{\mu}_k\varphi_r = 1$ for some $k\in\mathcal{I}$ and $r = 1, \hdots, p$. Since $\varphi_r = 1$, this implies that $\exists k: \bar{\mu}_k = 1$, i.e. for some nonzero $\vect{r}\in\mathbb{R}^n$ there holds $(\mat{B}+\mat{I}_n)^{-1}(\mat{I}_n-\mathcal{L})\vect{r} = \vect{r}$ or equivalently $(\mat{B}+\mathcal{L})\vect{r}  = \vect{0}_n$.
In other words, $\mat{B}+\mathcal{L}$ must be singular if $\mat{N}$ has an eigenvalue at 1. With $\mathcal{L}$ expressed in lower block triangular form as in \eqref{eq:laplacian_ordered}, the arguments below \eqref{eq:laplacian_ordered} establish that $\mathcal{L}_{11}+\mbox{diag}[b_1, \hdots, b_r]$ and $\mathcal{L}_{22} + \mbox{diag}[b_{r+1}, \hdots, b_n]$ are separately nonsingular, because $b_i \geq 0 \forall\, i \in \mathcal{I}$ and $\exists j \in \{1, \hdots, r\} : b_j > 0$. Thus, $\mat{B}+\mathcal{L}$ is nonsingular. It follows that $\mat{N}$ does not have an eigenvalue at $1$, and thus $\bar{\mat{M}}$ is invertible and Hurwitz. Lemma~\ref{lem:input_limit} establishes that final opinions are as given in \eqref{eq:x_final_stub}. \hfill$\square$

\begin{remark}
	Assumption~\ref{assm:L_C_invariant} places constraints on both the logic matrix $\mat{C}$ and entries of the network Laplacian $\mathcal{L}$ (note that $c_{kk} > 0$ simply implies that the $k^{th}$ topic is positively dependent on itself). The constraints are placed to ensure that if $\vect x(0) \in \mathcal{R}$, where $\mathcal{R}$ is defined in Lemma~\ref{lem:invar}, then $\mathcal{R}$ is an invariant set of \eqref{eq:compact_network_stubborn_02}. As it turns out, the same constraints are sufficient for convergence of Model 1, as identified in Theorems~\ref{thm:consensus_assm2} and \ref{thm:stub_assm2}. Generally speaking, if Assumption~\ref{assm:L_C_invariant} does not hold, then Model 1 may still converge, even though the desired invariance property may no longer hold. 
\end{remark}

\begin{remark}\label{rem:oblivious}
In \cite{parsegov2017_multiissue}, individual $i$ is said to be \textbf{oblivious} if $b_i = 0$ and $\nexists j \in \mathcal{I}$ such that $b_j > 0$ and there is directed path from $v_j$ to $v_i$. That is, an oblivious individual is not stubborn, and is not influenced by a stubborn individual's opinion via a directed path. Then, the sufficient condition involving $b_i$ in Theorem~\ref{thm:stub_assm2} can be interpreted as ensuring that there are no oblivious individuals.
\end{remark}

\begin{remark}
When $\mat{C} = \mat{I}_d$, the eigenvalue of $\mathcal{L}_{22}$ with the smallest real part, denoted $\lambda(\mathcal{L}_{22})$, governs the convergence rate of \eqref{eq:compact_network_stubborn_02}. Emerging results have studied how changes to network topology (including stubbornness $b_i$) impacts $\lambda(\mathcal{L}_{22})$ \cite{xia2017ground_laplacian,pirani2016grounded_laplacian}, but it is not clear how the introduction of $\mat{C} \neq \mat{I}_d$ affects such results, and may be a future research direction. 
\end{remark}

\section{Networks of Individuals with Model 2}\label{sec:model2}
Perhaps unsurprisingly, the lack of $\mat C$ in the first summand on the right of \eqref{eq:xi_update_C_stub_R} greatly simplifies the analysis, allowing for the establishing of a more comprehensive set of results, especially when stubborn individuals are present. We now present two theorems establish conditions for convergence in networks where individuals update according to \eqref{eq:xi_update_C_R}, and \eqref{eq:xi_update_C_stub_R}, respectively. Afterwards, we discuss the results on Model 1 and Model 2.
	
	\subsection{Analysis of Model 2}
	
	\begin{theorem}[No Stubborn Individuals]\label{thm:stability_continuous_C_R}
		Suppose that Assumption~\ref{assm:C_constant} holds for a given $\mat{C}$. Then for all $\vect x(0)$, with each individuals' opinions evolving according to \eqref{eq:xi_update_C_R}, the social network reaches a consensus on all topics exponentially fast if and only if $\mathcal{G}$ has a directed spanning tree, with limiting opinions
		\begin{equation}\label{eq:consensus_value_R}
		\lim_{t\to\infty} \vect{x}_i (t) = \Big(\sum_{r=1}^p \vect{\zeta}_r \vect{\xi}_r^\top \Big) \sum_{j=1}^n \gamma_j \m{x}_j(0), \;\forall\,i\in\mathcal{I}.
		\end{equation}
	\end{theorem}
	\emph{\textbf{Proof}:}
	The necessity of having a directed spanning tree is explained below Assumption~\ref{assm:graph}. For the proof of sufficiency, observe that convergence of \eqref{eq:3} depends on the eigenvalue properties of the matrix $\m{A} = \mc{L} \otimes \m{I}_d + \m{I}_n \otimes (\m{I}_d - \m{C})$. From \cite[Proposition 11.1.7]{bernstein2009matrixbook}, we obtain that the solution of \eqref{eq:3} is
	\begin{align} \label{eq:model}
	\m{x}(t) &= e^{-\mat A t} \m{x}(0) = \big(\mat N_1(t) \otimes \mat N_2(t) \big) \m{x}(0),
	\end{align}
	with matrix exponentials $\m{N}_1(t) = e^{-\mc{L}t}$ and $\m{N}_2(t) = e^{-(\m{I}_d - \m{C})t}$. Since $\mc{G}$ has a directed spanning tree, we know that $\lim_{t \to \infty} \m{N}_1(t) = \m{1}_n \vect{\gamma}^\top$, where $\m \gamma$ is defined in Lemma~\ref{lem:DST_graph}.
	Under Assumption 1, one can obtain 
	$\lim_{t \to \infty} \m{N}_2(t) = \sum_{r=1}^p \vect{\zeta}_r \vect{\xi}_r^\top \triangleq \mat P$. 
	It follows that
	\begin{align*}
	\lim_{t \to \infty} \m{x}(t) = \big((\m{1}_n \vect{\gamma}^\top) \otimes \mat P\big) \m{x}(0) 
	= \m{1}_n \otimes \Big(\Big(\vect{\gamma}^\top \otimes \mat P\Big)\m{x}(0)\Big).
	\end{align*}
	Thus, the final opinions of each individual are at the consensus value of $(\vect{\gamma}^\top \otimes \mat P )\m{x}(0) = \mat P \sum_{i=1}^n \gamma_i \m{x}_i(0)$.  \hfill$\square$
	
	
	
\begin{theorem}[Stubborn Individuals]\label{thm:stub_R}
	Suppose that Assumptions~\ref{assm:C_constant} and \ref{assm:graph} hold for given $\mat{C}$ and $\mathcal{G}=(\mathcal{V}, \mathcal{E}, \mathcal{A})$. If each individual's opinions evolve according to \eqref{eq:xi_update_C_stub_R}, $b_i\geq 0\,\forall\,i\in\mathcal{I}$, and $\exists j\in \{1, \hdots, r\} : b_j > 0$, then the system \eqref{eq:compact_network_stubborn_02} converges exponentially fast to
	\begin{align}\label{eq:final_stubborn_R}
	\lim_{t\to\infty} \vect x(t)&  = \Big((\mc{L}+\m{B})\otimes \mat I_d + \mat I_n \otimes (\m{I}_d - \m{C})\Big)^{-1}\nonumber \\
	&  \qquad \qquad \times(\m{B} \otimes \m{I}_{d}) \m{x}(0).
	\end{align}
\end{theorem}\vspace{-11pt}
\emph{\textbf{Proof}:}
With $\mathcal{L}$ expressed in lower block triangular form as in \eqref{eq:laplacian_ordered}, the arguments below \eqref{eq:laplacian_ordered} establish that all eigenvalues of $\mathcal{L}_{11}+\mbox{diag}[b_1, \hdots, b_r]$ and $\mathcal{L}_{22} + \mbox{diag}[b_{r+1}, \hdots, b_n]$ have strictly positive real part, because $b_i\geq 0\,\forall\,i\in\mathcal{I}$ and $\exists j \in \{1, \hdots, r\} : b_j > 0$. Thus, $-(\mathcal{L}+\mat{B})$ is Hurwitz. Let $\m M = (\mc{L}+\m{B})\otimes \mat I_d + \mat I_n \otimes (\m{I}_d - \m{C})$. 
From \cite[Proposition 7.2.3]{bernstein2009matrixbook}, we have that $\lambda_i = \mu_j + \rho_k$, where $\lambda_i, i \in \{1, \hdots nd\}$, $\mu_j, j \in \mathcal{I}$ and $\rho_k, k \in \mathcal{J}$ are eigenvalues of $\m M$, $\mc{L}+\m{B}$, and $(\m{I}_d - \m{C})$, respectively. Since $-(\mc{L}+\m{B})$ is Hurwitz and $-(\m{I}_d - \m{C})$ has $p$ zero eigenvalues and all other eigenvalues have negative real part, it follows that $-\m M$ is Hurwitz. Convergence to \eqref{eq:final_stubborn_R} is concluded from Lemma~\ref{lem:input_limit}. 
	\hfill $\qed$

\subsection{Discussions on Model 1 and Model 2}\label{ssec:discuss}

To place the focus on the effects of the logic matrix $\mat C$, let us compare networks with Model 1 and Model 2 having no stubborn individuals, i.e. \eqref{eq:x_system_continuous_C_02} and \eqref{eq:3}. Notice that if \eqref{cond:stable_C_02} is satisfied, then the consensus value is the same for both \eqref{eq:x_system_continuous_C_02} and \eqref{eq:3}, as recorded in \eqref{eq:stability_continuous_C_solution} and \eqref{eq:consensus_value_R}. Interestingly, this implies that \textit{if a consensus is reached}, then the presence of the matrix $\mat C$ in the interpersonal influence term has no effect on the limiting opinion values. In other words, if \eqref{cond:stable_C_02} is satisfied, then it does not matter whether individuals exchange opinions that have or have not been first assimilated using the logic matrix $\mat C$. However, \eqref{cond:stable_C_02} may not always be satisfied, implying that Model 1 is unstable, even if the conditions for Model 2 to be stable are fulfilled.
	
	It is clear when comparing Theorem~\ref{thm:stability_continuous_C} with Theorem~\ref{thm:stability_continuous_C_R} that the necessary and sufficient condition for consensus is much simpler for Model 2 compared with Model 1. Specifically, in Model 2, the only requirement on $\mat C$ and $\mathcal{G}$ are \textbf{separately} Assumption~\ref{assm:C_constant} and Assumption~\ref{assm:graph} ($\mathcal{G}$ has a directed spanning tree). This is similar to the discrete-time model in \cite{parsegov2017_multiissue}, where consensus (with no stubborn individuals) is reached by the system $\vect x(k+1) = (\mat W \otimes \mat D)\vect x(k)$ if and only if $\lim_{k\to\infty} \mat{D}^k$ exists and either (i) $\lim_{k\to\infty} \mat{D}^k = \mat 0_{d\times d}$, or (ii) $\lim_{k\to\infty} \mat{W}^k = \vect{1}_n\vect{v}^\top$ for some nonnegative vector $\vect{v}$. Here, $\mat D$ and $\mat{W}$ is the discrete-time counterpart to $\mat C$ and $\mathcal{L}$, respectively. In contrast, for Model 1, the necessary and sufficient condition \eqref{cond:stable_C_02} clearly depends on the combination of the network topology as encoded by $\mathcal{L}$, and the logical interdependence as encoded by $\mat{C}$. Given the same $\mat{C}$, two different $\mathcal{G}_1$ and $\mathcal{G}_2$ may have different convergence properties.
	
	Without logical interdependence, i.e. $\mat C = \mat I_d$, then the consensus value is $\sum_{j=1}^n \gamma_j \m{x}_j(0)$. Defining the projection matrix $\mat Y = \sum_{r=1}^p \vect{\zeta}_r \vect{\xi}_r^\top$, we see that the effect of the logical interdependence is to project $\sum_{j=1}^n \gamma_j \m{x}_j(0)$ onto the range space of $\mat Y$, which is equivalent to the nullspace of $\mat I_d - \mat C$. Put another way, the final consensus value in \eqref{eq:stability_continuous_C_solution} or \eqref{eq:3} is consistent with the logical interdependence structure, being a fixed point of $\mat C$. Recent work on the discrete-time model \cite{ye2018_CDC_logic} indicates that heterogeneity of $\mat C$ may create disagreement in the final opinions, even without stubborn individuals. It would be of great interest to investigate whether the same phenomena holds in either Model 1 or Model 2 (or both).

	Concerning Model 1, Corollaries~\ref{cor:C_constant_alpha} and~\ref{cor:C_constant_stable} show that for any given $\mat{C}$, it is always possible to reach a consensus if there is a sufficiently slow exchange of opinions (weights $a_{ij}$ are small). Sets of topics whose $\mat{C}$ have large $e_k$ and $\theta_k$ close to $\pi/2$ (as defined in Corollary~\ref{cor:C_constant_stable}) reflect a cognitive process \eqref{eq:xi_cognitiveprocess} where the opinions oscillate heavily and rapidly before settling to a consistent belief system. In such instances, the bound \eqref{eq:cor_C_constant_if} becomes  $\bar{l} < \frac{\vert 1 - \vert \lambda_k\vert \cos(\theta_k)\vert (1+\cos(\theta_k))}{\vert \lambda_k\vert \sin^2(\theta_k)} = \frac{(1-d_k)(1+\cos(\theta_k))}{e_k \sin(\theta_k)} < 0.5$. Thus, one method of guaranteeing consensus when individuals exchange assimilated opinions $\mat Cx_i(t) - \mat Cx_j(t)$ is to reduce the rate of interaction, by decreasing $a_{ij}$ to satisfy \eqref{eq:cor_C_constant_if}. On the other hand, rapid discussions \textit{may} lead to instability. We stress the word \textit{may}, because sometimes there is no risk. For instance, if the eigenvalues of $\mat C$ are all nonnegative real, then Model 1 will always consensus for any magnitude of the $a_{ij}$s. A simple example is when $\mat C$ is lower triangular (which reflects a cascade logic structure) and satisfies Assumption~\ref{assm:C_constant}. When topics are uncoupled, $\mat{C} = \mat{I}_d$, or the logical interdependences have not been assimilated into the exchanged opinions (Model 2), there is never such risk.
	
	In psychology and organisational science, it has been observed that an individual receiving external information at a high density that overwhelms their internal capacity to process the information can experience \emph{cognitive overload}, leading to a decrease in decision making abilities and response time to stimulus \cite{eppler2004information_overload,fox2007cognitive_overload}. When considering the above discussion, it appears that the instability that can arise in Model 1 when increasing exchange of assimilated opinions resembles behaviour similar to cognitive overload. More precisely, one interpretation of the results on Model 1 is that when individual $i$'s opinions $\vect x_i$ evolves by both an internal cognitive process and interpersonal social influence (the first and second term of \eqref{eq:xi_update_C} respectively), the internal process should be not ``overwhelmed" by exogenous social influence because individuals are assimilating logical interdependences into the opinions being exchanged, as captured by the term $\sum_{j\in\mathcal{N}_i} a_{ij} (\mat Cx_i(t) - \mat Cx_j(t))$. 
	
	One may wish to model the interpersonal influence process and introspective process as having different time-scales. Assuming $b_i = 0$ for all $i$ for simplicity of exposition, one possible approach is to multiply the first and second terms of \eqref{eq:xi_update_C_stub} with positive scalars $\alpha$ and $\beta$, respectively, with the relative magnitudes of $\alpha$ and $\beta$ capturing the relative time-scales of the two processes. The same can be done for Model 2, in \eqref{eq:xi_update_C_stub_R}. The clear distinguishing between assimilation of logical interdependence into the opinions, and the introspective process, as detailed in Remark~\ref{rem:assim}, enables one to model relative time-scales as suggested. The analysis above indicates that for Model 2, the convergence condition in Theorems~\ref{thm:stability_continuous_C_R} continues to apply irrespective of $\alpha$ and $\beta$. In contrast, it is possible that a fast interpersonal influence process, viz. large $\alpha$, can yield instability for Model 1 (see Theorem~\ref{thm:stability_continuous_C}, and Corollary~\ref{cor:C_constant_alpha}). We illustrate this with simulations below, and leave further investigations to future work.

\subsubsection{More Comments on Corollary~\ref{cor:C_constant_stable}}\label{sssec:discuss_v2}
\begin{figure}
	\centering
	\includegraphics[height=0.85\linewidth,angle=-90]{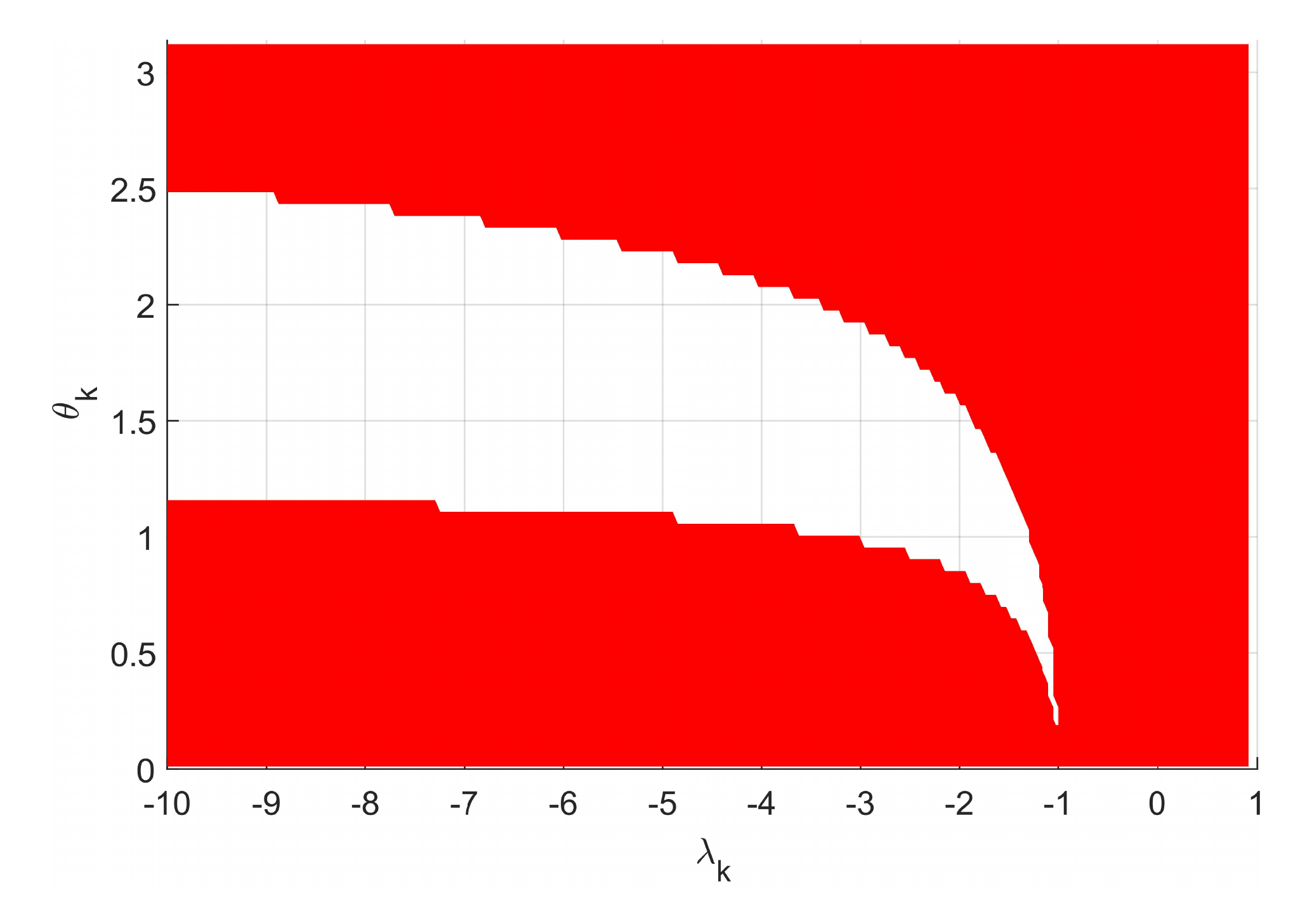}
	\caption{A plot of the switching of the bound $\bar{l} < \min \left\{\frac{\vert 1 - \vert \lambda_k\vert \cos(\theta_k)\vert (1+\cos(\theta_k))}{\vert \lambda_k\vert \sin^2(\theta_k)}, 0.5 \right\}$. The shaded red region indicates that $\frac{\vert 1 - \vert \lambda_k\vert \cos(\theta_k)\vert (1+\cos(\theta_k))}{\vert \lambda_k\vert \sin^2(\theta_k)} > 0.5$ and thus one requires $\bar{l} < 0.5$, while the white region indicates $\frac{\vert 1 - \vert \lambda_k\vert \cos(\theta_k)\vert (1+\cos(\theta_k))}{\vert \lambda_k\vert \sin^2(\theta_k)} < 0.5$.}
	\label{fig:bound_simulation}
\end{figure}

Now, we provide a simulation that helps to illustrate the switching of the bound between the two terms on the right hand side of \eqref{eq:cor_C_constant_if}. Fig.~\ref{fig:bound_simulation} shows the values of the parameter pair $\theta_k, \lambda_k$ for which one requires $\bar{l} < 0.5$ (red shaded region) or $\bar{l} < \frac{\vert 1 - \vert \lambda_k\vert \cos(\theta_k)\vert (1+\cos(\theta_k))}{\vert \lambda_k\vert \sin^2(\theta_k)}$ (white region) to guarantee that a consensus of opinions is achieved. One can immediately notice that we typically require $\bar{l} < \frac{\vert 1 - \vert \lambda_k\vert \cos(\theta_k)\vert (1+\cos(\theta_k))}{\vert \lambda_k\vert \sin^2(\theta_k)}$ for values of $\theta_k \approx \pi/2$ and $\lambda_k$ that are large and negative. Such $\theta_k, \lambda_k$ values correspond to systems of \eqref{eq:xi_cognitiveprocess} whose trajectories oscillate heavily before converging to a steady state. This is precisely what we identify in Section~\ref{ssec:discuss}; such scenarios are where one requires interpersonal interactions to be significantly weaker than the internal cognitive process.

A related, but different question is to determine when the bound \eqref{eq:cor_C_constant_if} is tight. This is a substantially more challenging question to answer, one which may be an interesting future research direction. We provide here an observation only; the bound $\bar{l} < 0.5$ is tight if the eigenvalues of $\mat{C}$ are all real (and satisfy Assumption~\ref{assm:C_constant}). The proof of this is found in the proof of Corollary~\ref{cor:C_constant_stable}. Triangular $\mat{C}$ are one class of logic matrices which where all the eigenvalues are all real. Despite the mathematical restrictiveness, triangular $\mat{C}$ may be more common in the context of this model, since such logic matrices describe interdependences between topics built from a single axiomatic statement \cite{ye2020_LogicTAC}.

\section{Simulations}\label{sec:simulations}
We now present a short numerical computation to illustrate some of our key results. The network $\mathcal{G}$ has a directed spanning tree, and the associated Laplacian $\mathcal{L}$  is 
\begin{equation}\label{eq:simulation_laplacian}
\mathcal{L} = \left[
\begin{array}{cccc:cccc}
1 & 0 & -1 & 0 & 0 & 0 & 0 & 0 \\
-1 & 1 & 0 & 0 & 0 & 0 & 0 & 0 \\
0 & -0.8 & 1 & -0.2 & 0 & 0 & 0 & 0 \\
0 & 0 & -1 & 1 & 0 & 0 & 0 & 0 \\ \hdashline
0 & 0 & 0 & -0.4 & 1 & 0 & -0.6 & 0 \\
0 & 0 & -0.2 & 0 & -0.8 & 1 & 0 & 0 \\
0 & 0 & 0 & 0 & 0 & -1 & 1 & 0 \\
-0.3 & -0.7 & 0 & 0 & 0 & 0 & 0 & 1 \\
\end{array}
\right].
\end{equation} 
The coupling matrix is given by 
\begin{equation}\label{eq:simulation_C}
\mat{C} = \begin{bmatrix} 1 & 0 & 0\\ -0.1 & 0.2 & 0.7 \\ 0.1 & -0.8 & 0.1  \end{bmatrix}
\end{equation}
This might describe the following set of complex topics. Topic 1: North Korea has nuclear weapons capable of reaching the USA. Topic 2: As its ally, China will defend North Korea against an attack. Topic 3: The USA will use its nuclear arsenal to eliminate North Korea's nuclear strike ability. Verify that $\mathcal{G}$ contains a directed spanning tree, see also \eqref{eq:laplacian_ordered}, with $r = 4$. The initial conditions $\vect{x}(0)$ are generated from a uniform distribution in the interval $[-1,1]$. 
Initially, we assume $b_i = 0,\forall\,i\in \mathcal{I}$.

\begin{figure*}[!ht]
\begin{minipage}{0.425\linewidth}
	\centering
	\includegraphics[height=0.85\linewidth,angle=-90]{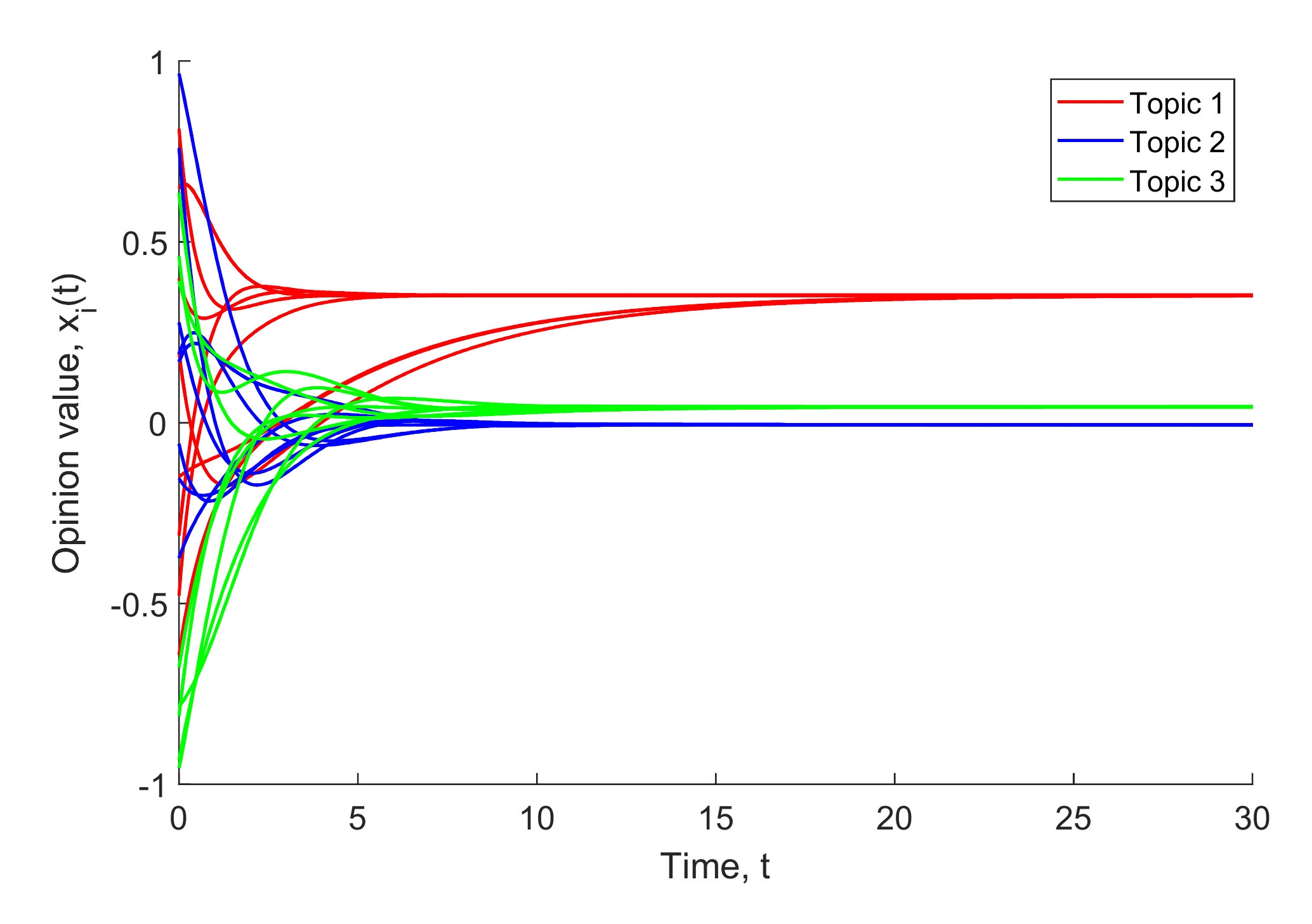}
	\caption{Model 1: Consensus is reached on all 3 topics when \eqref{cond:stable_C_02} in Theorem~\ref{thm:stability_continuous_C} is satisfied.}
	\label{fig:Model1_Stable}
\end{minipage}
\hfill
\begin{minipage}{0.425\linewidth}
	\centering
	\includegraphics[height=0.85\linewidth,angle=-90]{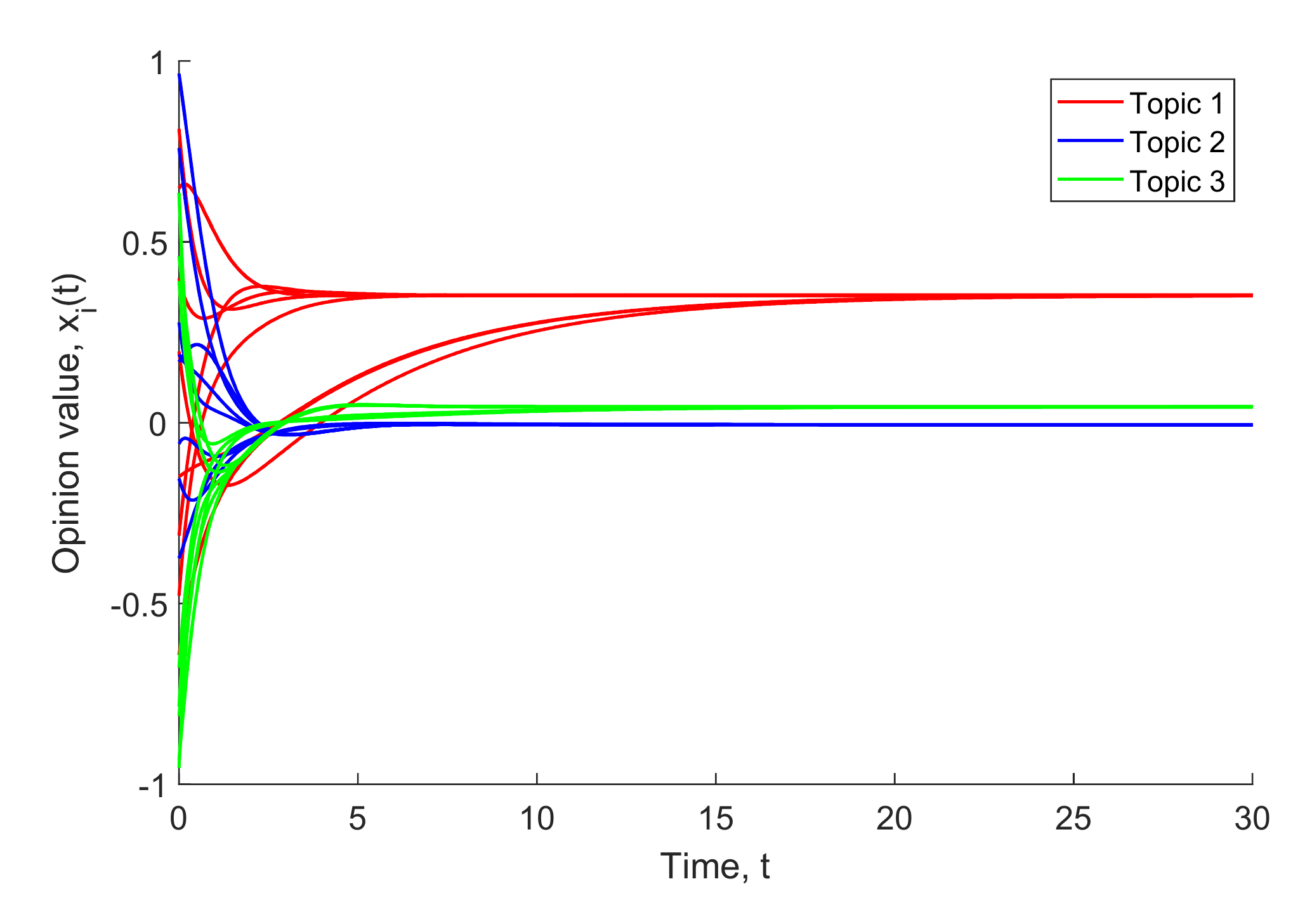}
	\caption{Model 2: Consensus is reached when $\mat C$ and $\mathcal{G}$ separately satisfy Assumption~\ref{assm:C_constant} and \ref{assm:graph}.}
	\label{fig:Model2_Stable}
\end{minipage}
\end{figure*}

\begin{figure*}[!ht]
\begin{minipage}{0.425\linewidth}
	\centering
	\includegraphics[height=0.85\linewidth,angle=-90]{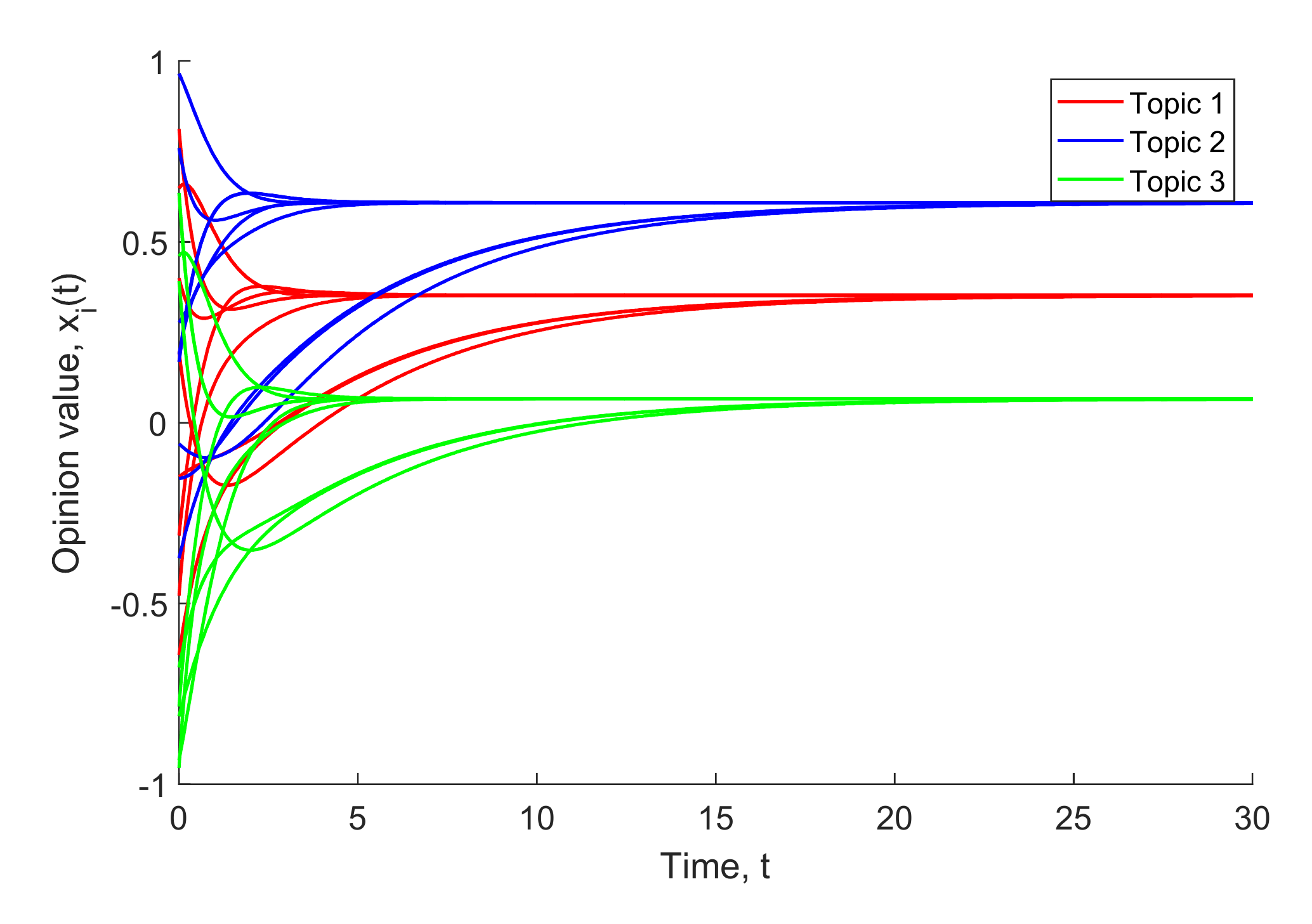}
	\caption{Model 1 and 2: When the topics are uncoupled, $\mat{C}= \mat{I}_d$, consensus is reached but the final consensus values are different due to the lack of logic coupling.}
	\label{fig:IdentityC}
\end{minipage}
\hfill
\begin{minipage}{0.425\linewidth}
	\centering
	\includegraphics[height=0.85\linewidth,angle=-90]{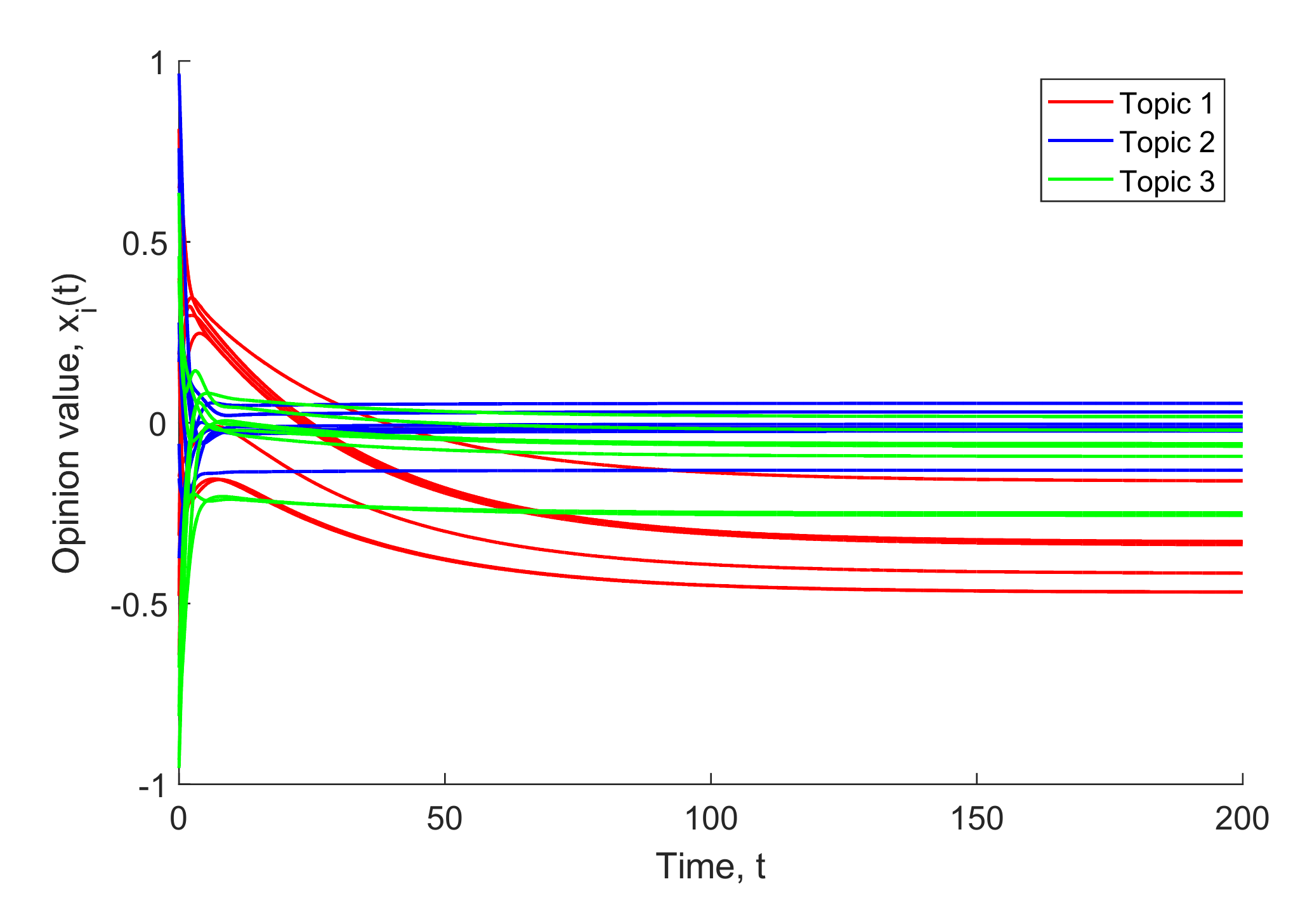}
	\caption{Model 1: In the presence of stubborn individuals, a state of persistent disagreement is achieved when the conditions in Theorem~\ref{thm:stub_assm2} are met.}
	\label{fig:Model1_Stubborn}
\end{minipage}
\end{figure*}

\begin{figure*}[!ht]
\begin{minipage}{0.425\linewidth}
	\centering
	\includegraphics[height=0.85\linewidth,angle=-90]{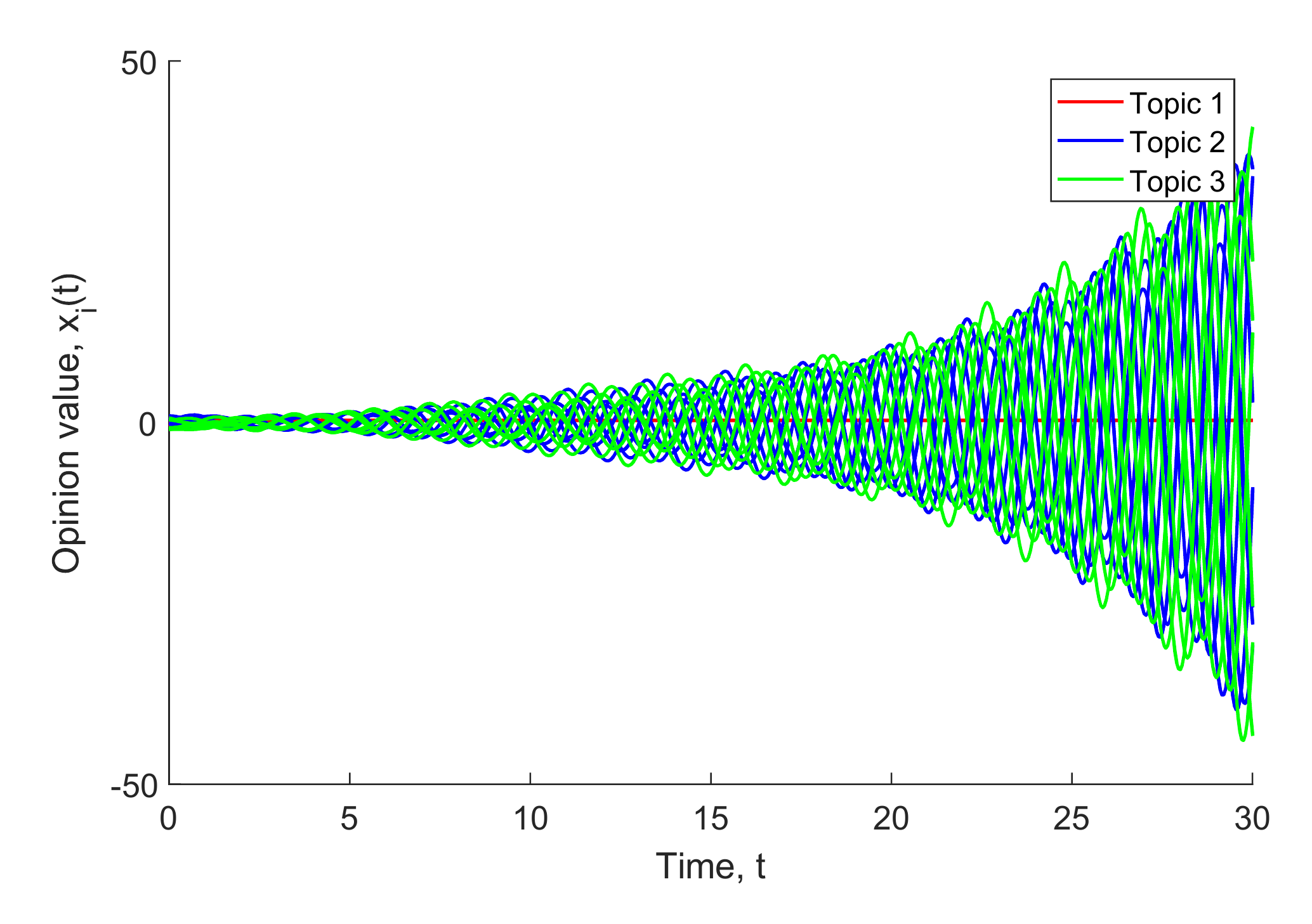}
	\caption{Model 1: Increasing the interpersonal interaction strengths means \eqref{cond:stable_C_02} of Theorem~\ref{thm:stability_continuous_C} is not satisfied, and the opinion system becomes unstable.}
	\label{fig:Model1_UnStable}
\end{minipage}
\hfill
\begin{minipage}{0.425\linewidth}
	\centering
	\includegraphics[height=0.85\linewidth,angle=-90]{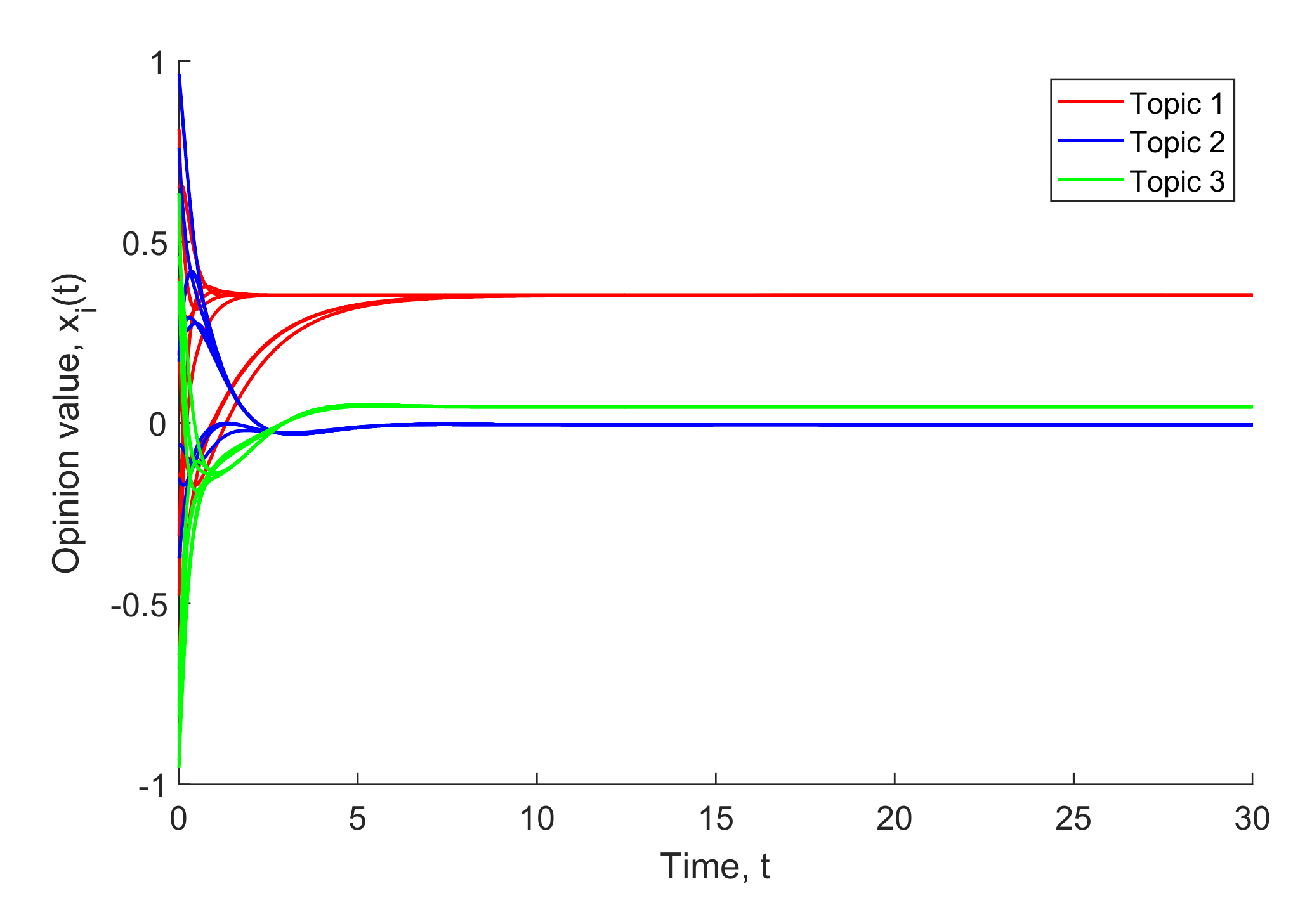}
	\caption{Model 2: Increasing the interpersonal interaction strengths changes the transient, but consensus is still achieved.}
	\label{fig:Model2_Stable_Fast}
\end{minipage}
\end{figure*}

The given $\mathcal{L}$ and $\mat{C}$ satisfy Assumptions~\ref{assm:C_constant}, \ref{assm:graph} and \ref{assm:L_C_invariant}, and also satisfies the condition of \eqref{cond:stable_C_02} in Theorem~\ref{thm:stability_continuous_C}. The dynamics of Model 1 and Model 2 are shown in Fig.~\ref{fig:Model1_Stable} and \ref{fig:Model2_Stable}, respectively. We see that the opinions for all 3 topics reach a consensus, and as discussed in Section~\ref{ssec:discuss}, the final consensus value for both models is the same, though the transients differ. For the same $\mathcal{L}$ and $\vect{x}(0)$, Fig.~\ref{fig:IdentityC} shows the case where the topics are uncoupled with $\mat{C} = \mat{I}_d$. With $\mat{C}$ as in \eqref{eq:simulation_C},  Topic 2 is coupled to Topics 1 and 3 by a negative and positive weight respectively. The coupling effect is clear: the consensus value of Topic 2 in Fig.~\ref{fig:Model1_Stable} is further from the consensus value of Topic 1 and closer to the consensus value of Topic 3 when compared to Fig.~\ref{fig:IdentityC}. Next, we introduce stubbornness, with $\vect{b} = [0, 0.1, 0, 0.05, 0, 0.4, 0, 0.3]^\top$, and the same $\vect{x}(0)$, $\mathcal{L}$, and $\mat{C}$ as above. Consistent with Theorem~\ref{thm:stub_assm2}, opinions converge to a persistent disagreement as shown in Fig.~\ref{fig:Model1_Stubborn}. Last, we return to $b_i = 0,\forall\,i\in \mathcal{I}$, and use the same $\vect{x}(0)$ and $\mat{C}$ (as in \eqref{eq:simulation_C}), but each edge weight is multiplied by 3, i.e. the new Laplacian $\bar{\mathcal{L}}$ satisfies $\bar{\mathcal{L}} = 3\mathcal{L}$. Eq. \eqref{cond:stable_C_02} of Theorem~\ref{thm:stability_continuous_C} is not satisfied; we see from Fig.~\ref{fig:Model1_UnStable} that the opinions diverge for Model 1, but consensus is still achieved for Model 2, Fig.~\ref{fig:Model2_Stable_Fast}.  

\section{Conclusions}\label{sec:con}
In this paper, we have proposed two continuous-time opinion dynamics model for a social network discussing opinions on multiple logically interdependent topics. When there are no stubborn individuals in the network, separate necessary and sufficient conditions are derived for networks to achieve a consensus of opinions in both models. The condition for Model 1 depends on the interplay between the logic coupling matrix and the graph topology, which is in contrast to Model 2, where separate conditions on the logic matrix and graph Laplacian matrix need to be satisfied. Further sufficient conditions for consensus were obtained for Model 1, to better understand the role of the logic matrix and graph topology. Networks with stubborn individuals were studied for both models, with sufficient conditions obtained for the opinions to converge to a limit. Future work will involve further analysis of networks with stubborn individuals in Model 1, and to consider heterogeneous logic matrices among the individuals. 

\begin{ack}                               
	We thank the Associate Editor and anonymous reviewers for
	their valuable comments which greatly improved this paper. In particular, Model 2 was developed following detailed comments from an anonymous reviewer. 
	
	M. Ye is supported in part by the European Research Council (ERC-CoG-771687) and the Netherlands Organization for Scientific Research (NWO-vidi-14134). B. D. O. Anderson is supported by the Australian Research Council (ARC) under grant \mbox{DP-160104500}, also by Data61-CSIRO. M. H. Trinh and H.-S. Ahn were supported by the National Research Foundation (NRF) of Korea under the grant NRF-2017R1A2B3007034
\end{ack}

\bibliographystyle{IEEEtran}        
\bibliography{MYE_ANU}           

\end{document}